\documentclass[12pt]{article}
\usepackage[dvips]{color,graphicx}
\usepackage{hyperref}
\usepackage{amsmath,amssymb}
\DeclareGraphicsExtensions{.eps}

\newcommand{\ap}{\ensuremath{\alpha'}} 
\newcommand{\ls}{\ensuremath{l_s}} 

\def\eps{\ensuremath{\epsilon}}
\def\del{\nabla}

\def\p{\partial}

\newcommand{\tr}{\mathop{\rm Tr}}

\def\expec#1{\langle #1 \rangle}
\def\ket#1{| #1 \rangle}
\def\bra#1{\langle  #1 |}

\def\slash#1{\ensuremath{\;/\!\!\!\! #1}}

\newcommand{\cF}{{\mathcal{F}}}

\newcommand{\cL}{\mathcal{L}}

\newcommand{\cN}{{\mathcal{N}}}
\newcommand{\cO}{{\mathcal{O}}}

\newcommand{\bR}{{\mathbf{R}}}
\newcommand{\bS}{{\mathbf{S}}}

\newcommand{\tret}{{t_{\mbox{\scriptsize ret}}}}
\newcommand{\xperp}{{\vec{x}_{\perp}}}
\newcommand{\tp}{{t^{\prime}}}
\newcommand{\xp}{{x^{\prime}}}
\newcommand{\zp}{{z^{\prime}}}
\newcommand{\Xp}{{X^{\prime}}}
\newcommand{\xperpp}{{\vec{x}_{\perp}^{\,\prime}}}
\newcommand{\tpp}{{t^{\prime\prime}}}
\newcommand{\xpp}{{x^{\prime\prime}}}
\newcommand{\zpp}{{z^{\prime\prime}}}
\newcommand{\Xpp}{{X^{\prime\prime}}}
\newcommand{\xperppp}{{\vec{x}_{\perp}^{\,\prime\prime}}}
\newcommand{\Xperppp}{{\vec{X}_{\perp}^{\,\prime\prime}}}
\newcommand{\zt}{{\tilde{z}}}

\begin{document}

\begin{titlepage}

\begin{flushright}
UTTG-03-10
\end{flushright}

\begin{center} \Large \bf Quantum Fluctuations and the 
Unruh Effect \\
in Strongly-Coupled 
Conformal Field Theories
\end{center}

\begin{center}
Elena C\'aceres$^{\dagger}$\footnote{elenac@ucol.mx}, Mariano Chernicoff$^{\star}$\footnote{mchernicoff@ub.edu},
Alberto G\"uijosa$^{\natural}$\footnote{alberto@nucleares.unam.mx}
and Juan F.~Pedraza$^{\natural}$\footnote{juan.pedraza@nucleares.unam.mx}

\vspace{0.2cm}
${}^{\dagger}$  Facultad de Ciencias,\\ Universidad de Colima,\\
Bernal D\'{\i}az del Castillo 340, Col. Villas San Sebasti\'an,  \\
Colima, Colima 28045, M\'exico\\
 \vspace{0.2cm}
$^{\star}$ Departamento de F\'{\i}sica,
\\Facultad de Ciencias, \\ Universidad Nacional Aut\'onoma de
M\'exico,\\ M\'exico D.F. 04510, M\'exico\\
{\small and}\\
Departament de F\'\i sica Fonamental, \\
Universitat de Barcelona,\\
Marti i Franqu\`es 1, E-08028 Barcelona, Spain\\

 \vspace{0.2cm}
$^{\natural}$ Departamento de F\'{\i}sica de Altas Energ\'{\i}as,
\\Instituto de Ciencias Nucleares, \\ Universidad Nacional Aut\'onoma de
M\'exico,\\ Apdo. Postal 70-543, M\'exico D.F. 04510, M\'exico

\vspace{0.2cm}

\end{center}

{\bf Abstract:} Through the AdS/CFT correspondence, we study a uniformly accelerated
quark in the vacuum of strongly-coupled conformal field theories in various dimensions, and determine the resulting stochastic fluctuations of the quark trajectory.
{}From the perspective of an inertial observer, these are quantum fluctuations induced by the gluonic radiation emitted by the accelerated quark. {}From the point of view of the quark itself, they originate from the thermal medium predicted by the Unruh effect. We scrutinize the relation between these two descriptions in the gravity side of the correspondence, and show in particular that upon transforming the conformal field theory from Rindler space to the open Einstein universe, the acceleration horizon disappears from the boundary theory but is preserved in the bulk. This transformation allows us to directly connect our calculation of radiation-induced fluctuations in vacuum with the analysis by de Boer \emph{et al.} of the Brownian motion of a quark that is on average static within a thermal medium. Combining this same bulk transformation with previous results of Emparan, we are also able to compute the stress-energy tensor of the Unruh thermal medium.

\vspace{0.2in}

\smallskip
\end{titlepage}
\setcounter{footnote}{0}

\section{Introduction and Summary}

The radiation emitted by an accelerated charge inevitably backreacts on the charge. One effect, present already at the classical level, is a reaction force on the charge, that tends to damp its motion. But if the system is quantized,
 one additionally expects the emission of radiation to induce stochastic fluctuations of the charge's trajectory. In the context of a quantum Abelian gauge theory, the first effect has been explored in \cite{monizsharp,martin,rosenfelder,johnson}, and the second, in \cite{johnson,parentani}. The discovery of the gauge/gravity duality \cite{malda,gkp,w, magoo} has opened the possibility of extending the exploration to the previously uncharted terrain of strongly-coupled non-Abelian gauge theories.

The first step in this direction was taken recently in \cite{lorentzdirac,damping}, where it was shown that the duality allows a simple derivation of an equation of motion for a `composite' or `dressed' quark that correctly incorporates the effects of radiation damping. The result is a non-linear generalization of the classic (Abraham-)Lorentz-Dirac equation \cite{dirac} that is physically sensible and (unlike Lorentz-Dirac) has no self-accelerating or pre-accelerating solutions. The damping effect follows directly from the fact that, in the context of this duality, the quark corresponds to the endpoint of a string, whose body codifies the profile of the non-Abelian (near and radiation) fields sourced by the quark, and can thus act as an energy sink. In other words, the quark has a tail, and it is this tail that is responsible for the damping force.

This mechanism had been previously established in the computations of the drag force exerted on the quark by a thermal plasma, which is described in dual language in terms of a string living on a black hole geometry \cite{hkkky,gubser}. The analysis of \cite{lorentzdirac,damping} makes it clear that the damping effect is equally present in the gauge theory vacuum, i.e., irrespective of whether or not there is a  black hole in the dual geometry (although, of course, the detailed form of the damping force is different). On the other hand, energy loss via the string does turn out to be closely associated with the appearance of a \emph{worldsheet} horizon, as noticed initially in \cite{gubserqhat,ctqhat} at finite temperature and emphasized in \cite{dragtime} for the zero temperature case. This association has been further studied in \cite{dominguez,xiao,beuf,nolineonthehorizon}.

As is customary in the gauge/gravity setting, the calculation in \cite{lorentzdirac,damping} treated the quark as a heavy
particle coupled to the fully quantized gluonic ($+$ other gauge theory) field(s). In this paper we go beyond
the classical description of the particle and study the
quantum fluctuations of the quark trajectory induced by its
coupling to the
gluonic field.

While we expect the physics of interest to us to be present under rather general circumstances, for simplicity we will restrict attention to the anti-de Sitter (AdS) / conformal field theory (CFT) subcases of the gauge/gravity duality, with the CFT defined on  Minkowski spacetime of arbitrary dimension $d$.
We begin in Section \ref{basicsubsec} by recalling the basic entries of the AdS/CFT dictionary that are of interest to us, as well as the results (\ref{eom})-(\ref{radiationrate}) of \cite{lorentzdirac,damping} on radiation damping for a quark that accelerates in the CFT vacuum.
In Section \ref{uniformaccelerationsubsec} we specialize to the case of uniform acceleration, deriving the relevant classical string embedding (\ref{stringtrajectory}) as a particular instance of the general solution obtained in \cite{mikhailov}, and verifying that the induced metric on the worldsheet contains a black hole (as had been established previously in \cite{xiao,ppz}), a fact that plays a central role in our investigation. Such a black hole would in fact be present for any accelerated quark trajectory \cite{dragtime}, but we are able to carry out the calculations of interest to us only in the case where the worldsheet geometry is static, which corresponds to uniform acceleration.

When we go beyond the classical description of the string, two new effects are found, both of which are suppressed by a factor of the string length divided by the AdS curvature radius, or, equivalently (via (\ref{lambda})), by an inverse factor of the CFT coupling. On the one hand, we pick up the usual quantum fluctuations arising from the determinant of the path integral over string embeddings. These are present even for a static string (see, e.g., \cite{dorn,greensite}), and lead to spontaneous deviations from the average endpoint/quark trajectory of the type studied, e.g., in \cite{johnson2}. On the other hand,  the worldsheet black hole emits Hawking radiation, which populates the various modes of oscillation of the string. In what follows we will concentrate solely on this second effect, which is present only for an accelerated trajectory and is thus associated with the quantum fluctuations induced by the gluonic radiation emitted by the quark. Needless to say, in the future it would be interesting to determine the way in which the quark behavior is modified when both effects are combined.

As always, small perturbations about the average string embedding (\ref{stringtrajectory}) are described by free scalar fields propagating on the corresponding induced worldsheet geometry.  Our task is then to quantize these modes and establish the way in which the excitations generated in them by the presence of the worldsheet horizon make the string endpoint fluctuate. This analysis is in complete parallel with \cite{rangamani,sonteaney}, where the same question was studied for a static string on the (planar) Schwarzschild-AdS$_{d+1}$ geometry, which is dual to a static quark in a thermal bath of the CFT. It was shown in those works that worldsheet Hawking radiation indeed gives rise to the expected Brownian motion of the endpoint/quark, whose detailed form is captured by a generalized Langevin equation. The authors of \cite{rangamani} reached this conclusion in arbitrary dimension by assuming (following \cite{lm,ff}) that the state of the quantized embedding fields is the usual Hartle-Hawking (or Kruskal) vacuum, which describes the black hole in equilibrium with its own thermal radiation. The authors of \cite{sonteaney} focused on the case $d=4$ and followed a different but equivalent route, employing the dual relation between the Kruskal extension of the Schwarzschild-AdS geometry and the CFT Schwinger-Keldysh formalism \cite{maldaeternal,herzogson,ct}, together with the known connection between the latter and the generalized Langevin equation. These calculations were later generalized and elaborated on in \cite{iancu,giecold,casalderrey,deboer}.

When we attempt to run through either of these procedures for the case of a quark undergoing uniform acceleration at zero temperature, the analysis is complicated by the explicit time dependence present in the worldsheet geometry (\ref{wsmetric}), which is only to be expected, given that the velocity and rate of radiation of the quark vary as time marches on. It is then natural to expect the problem to simplify if instead of working in the coordinates appropriate for an inertial observer we transform to a Rindler coordinate system adapted to an observer sitting on the quark. We therefore postpone the study of the string fluctuations until Section \ref{fluctuationssec}, and dedicate Section \ref{unruhsec} to a close scrutiny of the relevant transformations and their physical consequences. This exercise turns out to be rather interesting in itself, and sheds light on the AdS implementation of the celebrated Unruh effect \cite{unruh} (for reviews, see, e.g., \cite{unruhreviews}). Earlier analyses of this implementation can be found in \cite{xiao,ppz}, as well as in the very recent work \cite{hirayama}, which appeared while the present paper was in preparation.

In Section \ref{rindlersubsec} we present the bulk diffeomorphism that implements the transition from Minkowski to Rindler coordinates (which we denote with primes), equation (\ref{rindlercoordsads}). This transformation gives rise to an acceleration horizon both in the boundary and bulk descriptions. As a result, a state that is pure from the inertial perspective will generally be mixed from the point of view of the Rindler observers, because the field degrees of freedom accessible to the latter will be entangled with degrees of freedom in the region beyond their horizon, which they must trace over. In particular, the pure AdS geometry expressed in Rindler coordinates, equation (\ref{rindlermetricads}), which is dual to the Minkowski vacuum of the CFT (as evidenced by the vanishing of the expectation value of the stress-energy tensor), is interpreted as a thermal bath at the expected Unruh temperature (\ref{unruhtemp}). In Rindler coordinates, the string embedding takes the form (\ref{stringtrajectoryp}), which as expected is static and bends towards the Rindler horizon.

In Section \ref{conformalsubsec} we observe that the Rindler horizon of the CFT can be removed via a Weyl transformation. The resulting geometry (\ref{openeinsteinmetriccft}) (which we label with double primes) is that of the open Einstein universe. Following \cite{theisen}, we identify the corresponding bulk transformation, equation (\ref{openeinsteincoordsads}). Unlike the diffeomorphism discussed in Section \ref{rindlersubsec}, it drastically alters the radial foliation of the AdS geometry. Even though, by construction, in this new conformal frame the acceleration horizon is no longer visible in the boundary description, we show that it is still present in the bulk, but lies at the fixed radial position that according to the AdS/CFT dictionary corresponds to the Unruh temperature (\ref{unruhtemp}). In other words, after the transformation (\ref{openeinsteincoordsads}), the thermal character of the CFT state arises not from entanglement with degrees of freedom that lie beyond a spacetime horizon, but from the direct identification of the specific energy scale (\ref{unruhtemp}) as the temperature of the CFT, in exact parallel with the dual interpretation of the Schwarzschild-AdS geometry.

{}From the doubly primed AdS metric (\ref{openeinsteinmetricads}), we can extract the expectation value of the stress-energy tensor in the conformal Minkowski vacuum of the CFT on the open Einstein geometry. The result, given in (\ref{tmunupp}), is in complete agreement with \cite{roberto}, where this same quantity was computed in the context of a more general investigation of hyperbolic black holes in AdS/CFT. It was elucidated long ago \cite{bunch,cdowker,bd} that for even $d$ this vev is shifted to a non-zero value  as a result of the Weyl anomaly, which implies that the transformation that takes us from Rindler to open Einstein spacetime is not a true symmetry of the CFT. The AdS counterpart of this statement is also well understood \cite{hs}. Using this information and the results obtained in \cite{roberto} for the conformal Rindler vacuum of the CFT on the open Einstein universe, we can translate back to Rindler spacetime to determine the energy-momentum tensor of the Unruh thermal medium, equation (\ref{tmunumediumresult}).

We close Section \ref{conformalsubsec} by noting that in the doubly primed coordinate system, the string  embedding (\ref{stringtrajectorypp}) is static and completely vertical. This encourages us to carry out our study of small perturbations about the average string trajectory precisely in this frame, and in Section \ref{fluctuationssec} we finally proceed to do so. Interestingly, both the base string embedding (\ref{stringtrajectorypp}) and the background geometry (\ref{openeinsteinmetricads}) at the location of the string are found to coincide exactly with the $d=2$ \emph{thermal} setup analyzed in \cite{rangamani}, which allows us to obtain the information we are after simply by translating to our language the results of that work. This close relation between the quantum fluctuations of the uniformly accelerated quark on Minkowski spacetime and the thermal fluctuations of a static quark in a thermal medium is evidently a direct consequence of the Unruh effect, but the reader should be aware that, for $d>2$, the detailed properties of this thermal medium are found to differ from those of the familiar homogeneous and isotropic thermal ensemble dual to the Schwarzschild-AdS$_{d+1}$ geometry.

In Section \ref{langevinunruhsubsec}, we carry over from \cite{rangamani} the generalized Langevin equation (\ref{langevinpp}), describing the way in which the thermal medium on the open Einstein geometry makes our quark fluctuate, and deduce its local approximation (\ref{locallangevinpp}), which is valid when the fluctuations are examined over time scales that are large compared to the quark Compton wavelength. In Section \ref{langevininertialsubsec}, we then translate all relevant quantities back to the original, Minkowski (unprimed) frame, thereby concluding that the radiation emitted by the quark induces quantum fluctuations of its trajectory that obey the equations (\ref{langevinlong})-(\ref{etakappa}), which constitute our main result. Within an appropriate range of temporal resolutions, these equations simplify to the local form (\ref{locallangevinlong})-(\ref{eta0kappa0}).

The most prominent feature of these equations (in either nonlocal or local form) is their manifest time dependence, which is in marked contrast with the static nature of the open Einstein thermal medium, but was of course entirely expected, given that, as we emphasized above, the uniformly accelerated quark plus gluonic radiation system is certainly not in equilibrium.   Another salient property is the anisotropy between the longitudinal and transverse directions, which is again inherent to the definition of the system in the inertial frame. In this second respect, our equations of motion for the fluctuations of a quark that undergoes uniform acceleration in the CFT vacuum are somewhat akin to those of \cite{iancu,casalderrey}, which considered thermal fluctuations of a quark ploughing at constant (and possibly relativistic) velocity through a thermal plasma. In our case, the anisotropy goes so far as to result in a longitudinal equation of motion that, unlike the transverse equation, contains a term linear in the fluctuation and is consequently not of generalized Langevin form. As discussed below (\ref{eta0kappa0}), another curious feature is found in the signs of this and the frictional term, which in the longitudinal case turn out to be counter-intuitive  within a certain period of time. Curiosity aside, the direct connection with the generalized Langevin equation in the open Einstein frame of course ensures that the novel longitudinal equation leads to physically sensible evolution. The fact that we have been able to get our hands on this strongly coupled physics constitutes yet another illustration of the usefulness of the AdS/CFT correspondence.

\section{AdS/CFT and Uniformly Accelerated Quarks}\label{acceleratedquarksec}

\subsection{Basic setup and radiation damping}\label{basicsubsec}

According to the AdS/CFT correspondence \cite{malda,gkp,w},
string theory (or M-theory) on a background that asymptotically approaches the $(d+1)$-dimensional anti-de Sitter (AdS) spacetime\footnote{Cross a compact space, which will be mentioned briefly around (\ref{lambda}) but will otherwise play no role in our discussion.}
\begin{equation}\label{metric}
ds^2_{\mathrm{AdS}}=G_{MN}dx^M dx^N={L^2\over z^2}\left(
-dt^2+d\vec{x}^{\,2}+{dz^2}\right)~,
\end{equation}
where $\vec{x}\equiv(x_1,\ldots,x_{d-1})$,
is dual to a $d$-dimensional conformal field theory (CFT).
The directions $x^{\mu}\equiv(t,\vec{x})$ are parallel to the AdS boundary $z=0$, and are
directly identified with the CFT directions.
 The
radial direction $z$ is mapped holographically into a variable length/energy scale in the CFT, in such a way that $z=0$ and $z\to\infty$ are respectively dual to the ultraviolet (UV) and infrared (IR) limits of the CFT \cite{uvir,bklt}.

When AdS is radially foliated with Poincar\'e (or horospheric) coordinates as in (\ref{metric}), the dual CFT lives on Minkowski spacetime, $ds^2_{\mathrm{CFT}}=\eta_{\mu\nu}dx^{\mu}dx^{\nu}$, but by choosing different foliations we can obtain gravity descriptions of the same CFT on other background geometries. Any asymptotically AdS metric can be written in the Fefferman-Graham \cite{fg} form
\begin{equation}\label{feffermangraham}
ds^2_{\mathrm{aAdS}}={L^2 \over z^2}\left(g_{\mu\nu}(z,x)dx^{\mu}dx^{\nu}+dz^2\right)~,
\end{equation}
from which the dual CFT metric $ds^2_{\mathrm{CFT}}=g_{\mu\nu}(x) dx^{\mu}dx^{\nu}$ can be directly read off as $g_{\mu\nu}(x)= g_{\mu\nu}(0,x)$. The full function $g_{\mu\nu}(z,x)$ is uniquely determined (via the Einstein equations in the bulk) from this boundary value together with data dual to the expectation value of the CFT stress-energy tensor $T_{\mu\nu}(x)$. More specifically, in terms of the near-boundary expansion
\begin{equation}\label{metricexpansion}
g_{\mu\nu}(z,x)=g_{\mu\nu}(x)+z^2 g^{(2)}_{\mu\nu}(x)+\ldots +z^d g^{(d)}_{\mu\nu}(x)+ z^d \log (z^2) h^{(d)}_{\mu\nu}(x)+\ldots~,
\end{equation}
the standard GKPW recipe for correlation functions \cite{gkp,w} leads after appropriate holographic renormalization to \cite{dhss,skenderis} (see also \cite{bk,myers,ejm})
\begin{equation}\label{graltmunu}
\expec{T_{\mu\nu}(x)}={ d\, L^{d-1} \over 16\pi G^{(d+1)}_{\mbox{\scriptsize N}}}\left(g^{(d)}_{\mu\nu}(x)+X^{(d)}_{\mu\nu}(x)\right)~,
\end{equation}
where $X^{(d)}_{\mu\nu}=0$ $\forall$ odd $d$,
\begin{eqnarray}\label{xmunu}
X^{(2)}_{\mu\nu}&=&-g_{\mu\nu}g^{(2)\alpha}_{\alpha}~,\\
X^{(4)}_{\mu\nu}&=&-{1\over 8}g_{\mu\nu}\left[\left(g_{\alpha}^{(2)\alpha}\right)^2
-g_{\alpha}^{(2)\beta}g_{\beta}^{(2)\alpha}\right]
-{1\over 2}g_{\mu}^{(2)\alpha}g_{\alpha\nu}^{(2)}
+{1\over 4}g^{(2)}_{\mu\nu}g_{\alpha}^{(2)\alpha}~,\nonumber
\end{eqnarray}
and $X^{(6)}_{\mu\nu}$ is given by a similar but more
longwinded
expression that we will not transcribe here. In (\ref{xmunu}) it is understood that the indices of the tensors
$g^{(n)}_{\mu\nu}(x)$ are raised with the inverse boundary metric $g^{\mu\nu}(x)$.

 Known examples of this duality include the cases $d=2,3,4,6$, which involve the near-horizon geometries and low-energy worldvolume theories associated with systems of multiple D1/D5-, M2-, D3- and M5-branes, respectively \cite{magoo}. For particle theory applications we are mainly interested in the case $d=4$, where the best understood example equates Type IIB string theory on AdS$_5\times\bS^5$ (with a constant dilaton and $N_c$ units of Ramond-Ramond five-form flux through the five-sphere) to
$\cN=4$ $SU(N_c)$ super-Yang-Mills (SYM) with 't~Hooft coupling
\begin{equation}\label{lambda}
\lambda\equiv g_{YM}^2 N_c={L^4\over \ls^4}.
\end{equation}
Replacing the five-sphere with other compact geometries one obtains gravity duals of CFTs with fewer supersymmetries.
For all $d$, we will find it notationally convenient to use the rightmost expression as a definition of $\lambda$, and to refer to the CFT fields as `gluonic'.

It follows trivially from (\ref{graltmunu}) that the string theory state described by the unperturbed metric (\ref{metric}) is dual to a CFT state where $\expec{T_{\mu\nu}(x)}=0$, i.e., the Minkowski vacuum. The closed string sector describing (small or large) fluctuations on top of this bulk geometry fully captures the gluonic physics. An additional, `quark' sector can be introduced into the CFT by appropriately adding to the bulk a set of $N_f$ `flavor' D-branes, whose excitations are described by open strings \cite{kk}. In this context, an isolated quark is dual to an open string that extends radially from the AdS horizon at $z\to\infty$ to a position $z=z_m$ where it ends on the flavor branes. In particular, a static, purely radial string corresponds to a static quark, and by computing the energy of the former one finds that $z_m$ is related to the quark mass $m$ through
\begin{equation}\label{zm}
z_m={\sqrt{\lambda}\over 2\pi m}~.
\end{equation}
We will work in the regime where the geometry is weakly curved in units of the string length, $L\gg\ls$, in which the string (or M-) theory in the bulk essentially reduces to supergravity, and the dual CFT is strongly coupled.
We also assume that $N_f\ll N_c$ and consequently neglect the backreaction of the flavor branes on the geometry; in the field theory this corresponds to working in a `quenched' approximation which disregards quark loops.

In more detail, it is really the $z=z_m$ endpoint of the string that is dual to the quark, while the body of the string encodes the (near and radiation) gluonic field profile set up by the quark. It should be borne in mind that the quark so described is automatically not `bare' but `composite' or `dressed' \cite{martinfsq,damping}, and is surrounded by a gluonic cloud with characteristic thickness (Compton wavelength) $z_m$.

 The string dynamics follows as usual from the Nambu-Goto action
\begin{equation}\label{nambugoto}
S_{\mbox{\scriptsize NG}}=-{1\over 2\pi\ap}\int
d^2\sigma\,\sqrt{-\det{h_{ab}}}\equiv \int
d^2\sigma\,\cL_{\mbox{\scriptsize NG}}~,
\end{equation}
where $h_{ab}\equiv\p_a X^M\p_b X^N G_{MN}(X)$ ($a,b=0,1$) denotes
the induced metric on the worldsheet.
The quark worldline is identified with the trajectory of the string endpoint, $x^{\mu}(\tau)=X^{\mu}(\tau,z_m)$.
We can exert an external force $\vec{F}$ on the  endpoint/quark by turning on an electric field $F_{0i}=F_i$ on the flavor branes. This amounts to adding to the Nambu-Goto action the
usual minimal coupling
\begin{equation}\label{externalforce}
S_{\mbox{\scriptsize F}}=\int
d\tau\,A_{\mu}(x(\tau))\p_{\tau}x^{\mu}(\tau)~.
\end{equation}

The string here is being described, as usual, in first-quantized
language, and, as long as it is sufficiently heavy, we are allowed to treat it semiclassically.
In CFT language, then,
we are at this point coupling a first-quantized quark to the strongly-coupled gluonic fields, and then carrying out the full path integral over the latter (the result of which is codified by the AdS spacetime), but treating the path integral over the quark trajectory $x^{\mu}(\tau)$ in a saddle-point approximation.

Variation of the string action $S_{\mbox{\scriptsize NG}}+S_{\mbox{\scriptsize F}}$ implies the standard Nambu-Goto equation of motion for all interior points of the string, plus the usual boundary condition \cite{acny}
\begin{equation}\label{stringbc}
\Pi^{z}_{\mu}(\tau)|_{z=z_m}=\cF_{\mu}(\tau)\quad\forall~\tau~,
\end{equation}
where
\begin{equation}\label{pizmu}
\Pi^{z}_{\mu}\equiv \frac{\p\cL_{\mbox{\scriptsize NG}}}{\p(\p_z X^{\mu})}
={\sqrt{\lambda}\over 2\pi}\left(\frac{(\p_{\tau}X)^2 \p_z X_{\mu}-(\p_{\tau}X\cdot \p_z X)\p_{\tau}X_{\mu}}{z^2\sqrt{(\p_{\tau}X\cdot \p_z X)^2-(\p_{\tau}X)^2(1+(\p_z X)^2)}}\right)
\end{equation}
is the worldsheet Noether current associated with spacetime momentum, and we have
defined $\cF_{\mu}\equiv-F_{\nu\mu}\p_{\tau}x^{\nu}$. The latter coincides with
the Lorentz $d$-force
$(-\gamma\vec{F}\cdot\vec{v},\gamma\vec{F})$ if the parameter $\tau$ is chosen to be the quark proper time, which  will be understood henceforth unless otherwise stated.

When the quark accelerates, it will radiate, and we generally expect this radiation to exert a damping force on the quark already at the classical level. A given quark/endpoint semiclassical trajectory $x^{\mu}(\tau)=X^{\mu}(\tau,z_m)$ is thus determined only through the combined effect of the applied external force and the concomitant backreaction of the gluonic fields.  In \cite{lorentzdirac,damping} it was shown, building upon the results of \cite{mikhailov,dragtime}, that the standard string dynamics, and in particular the boundary condition (\ref{stringbc}), imply that the quark obeys the equation of motion
\begin{equation}\label{eom}
{d\over d\tau}\left(\frac{m{d x^{\mu}\over d\tau}-{\sqrt{\lambda}\over 2\pi m} \cF^{\mu}}{\sqrt{1-{\lambda\over 4\pi^2 m^4}\cF^2}}\right)=\frac{\cF^{\mu}-{\sqrt{\lambda}\over 2\pi m^2} \cF^2 {d x^{\mu}\over d\tau}}{1-{\lambda\over 4\pi^2 m^4}\cF^2}~.
\end{equation}
This is a generalized, non-linear version of the Lorentz-Dirac equation, whose physical content is most clearly displayed when we rewrite it in the form
 \begin{equation}\label{eomsplit}
 {d P^{\mu}\over d\tau}\equiv {d p_q^{\mu}\over d\tau}+{d P^{\mu}_{\mbox{\scriptsize rad}}\over d\tau}=\cF^{\mu},
 \end{equation}
 where
 \begin{equation}\label{pq}
 p_q^{\mu}=\frac{m{d x^{\mu}\over d\tau}-{\sqrt{\lambda}\over 2\pi m} \cF^{\mu}}{\sqrt{1-{\lambda\over 4\pi^2 m^4}\cF^2}}
 \end{equation}
is the intrisic $d$-momentum of the quark (including the near-field contribution, and satisfying the on-shell relation $p_q^2=-m^2$), and
\begin{equation}\label{radiationrate}
{d P^{\mu}_{\mbox{\scriptsize rad}}\over d\tau}={\sqrt{\lambda}\, \cF^2 \over 2\pi m^2}\left(\frac{{d x^{\mu}\over d\tau}-{\sqrt{\lambda}\over 2\pi m^2} \cF^{\mu} }{1-{\lambda\over 4\pi^2 m^4}\cF^2}\right)
\end{equation}
is the rate at which $d$-momentum is carried away from the quark by gluonic radiation (which in the limit of infinite mass reduces to the standard Lienard formula from classical electrodynamics \cite{mikhailov}).

\subsection{Solution for uniform acceleration}\label{uniformaccelerationsubsec}

As announced in the Introduction, in the present paper we would like to go one step beyond the classical approximation for the quark,
and study quantum fluctuations $\delta x^{\mu}(\tau)$ induced on the quark trajectory by its coupling to the gluonic field. For this, in the gravity side of the AdS/CFT correspondence we need to identify the string configuration dual to the average quark trajectory of interest, and analyze small perturbations about it.

 Of course, specifying an endpoint trajectory $x^{\mu}(\tau)=X^{\mu}(\tau,z_m)$ does not uniquely determine the full evolution of the string, just like specifying the quark trajectory does not uniquely determine the gluonic field profile. Both on the AdS and CFT sides, for any given endpoint/quark worldline there exist an infinite number of configurations, which differ in the boundary condition on the string/gluonic waves at infinity (or equivalently, in the corresponding initial conditions). As in \cite{dragtime,lorentzdirac,damping}, we will focus solely on solutions that are \emph{retarded}, in the sense that the string/gluonic configuration at any given time depends only on the behavior of the string endpoint at \emph{earlier} times. These are the solutions that capture the physics of present interest, with influences propagating outward from the quark to infinity.

 Remarkably, the retarded solution to the Nambu-Goto equation of motion on AdS is known for an \emph{arbitrary} timelike trajectory of
the endpoint/quark \cite{mikhailov}.\footnote{The paper \cite{mikhailov} considered a string on AdS$_5$, but the resulting solution can be readily generalized to AdS$_{d+1}$ with other values of $d$.} In terms of the coordinates used in
(\ref{metric}), and directly parametrized in terms of the endpoint/quark worldline $x^{\mu}(\tau)$ and the external $d$-force $\cF_{\mu}(\tau)$ (related to $x^{\mu}(\tau)$ through (\ref{eom})), it reads \cite{dragtime,damping}
 \begin{equation}\label{mikhsolzm}
 X^{\mu}(\tau,z)=\left(z-z_m
\over{\sqrt{1-z^4_m \slash{\cF}^2}}\right)\left({dx^{\mu}\over
d\tau}-z^2_m\slash{\cF}^{\mu}\right)+x^{\mu}(\tau)~.
\end{equation}

Generically, a black hole is found to develop \emph{on the string worldsheet} described by this solution \cite{dragtime}, even though no \emph{spacetime} black hole is present in the background geometry (\ref{metric}). The quantum fluctuations we are after will turn out to be induced by the Hawking radiation emerging from the horizon of this worldsheet black hole, in complete parallel with the thermal fluctuations determined in the analyses of the CFT at finite temperature \cite{rangamani,sonteaney,iancu,casalderrey,deboer} (indeed, this parallel had already been noted briefly in the Introduction of \cite{iancu}). It would be interesting but rather difficult to determine the rate of Hawking radiation for the generically dynamical black hole associated with an arbitrary quark worldline, so for simplicity we will specialize to the case of uniform acceleration, where the worldsheet black hole will turn out to be static (for the same reason, the studies of thermal fluctuations  have been thus far restricted to the case where the quark is static \cite{rangamani,sonteaney,deboer} or moving with constant velocity \cite{iancu,casalderrey} with respect to the thermal medium).

Henceforth we will assume that the (average) motion of the quark is purely along direction $x\equiv x^1$, and results from the application of a constant external force $\vec{F}=(F,0,\ldots)$, with $F>0$. In this case
the  equation of motion (\ref{eom}) implies that the quark moves with constant \emph{proper} acceleration\footnote{Note that when we turn on the constant electric field needed to accelerate the quark, the embedding of the flavor branes in AdS changes \cite{kundu}, implying in turn a modification of the relation between the radial location $z_m$ of the string endpoint and the Lagrangian mass of the quark. As a result, $m$ in (\ref{zm}) and (\ref{A}) must be interpreted as the effective quark mass in the presence of the electric field. In the case where the quark is subjected to an arbitrary time-dependent external force, it would be much more complicated to work out the detailed relation between the corresponding masses, but the same general idea still applies.  We are grateful to Arnab Kundu for a discussion of this point.} \cite{damping}
 \begin{equation}\label{A}
 A\equiv \sqrt{{d^2 x^{\mu}\over d\tau^2}{d^2 x_{\mu}\over d\tau^2}}={d\over dt}(\gamma
 v)=\frac{F}{\sqrt{m^2-{\lambda F^2\over 4\pi^2 m^2}}}~,
 \end{equation}
and so follows the hyperbolic trajectory
\begin{equation}\label{quarktrajectory}
x(t)=\sqrt{A^{-2}+t^2}~,
\end{equation}
where for convenience we have made a particular choice of the spacetime origin.
The proper time of the quark is then
\begin{equation}\label{tau}
\tau=A^{-1}\mathrm{arcsinh}(At)~.
\end{equation}
Using these data as input, and defining $\tret\equiv t(\tau,z_m)$, the general solution (\ref{mikhsolzm}) takes the form
\begin{eqnarray}\label{tretsol}
t(\tret,z)&=&(z-z_m)\gamma(\tret)\left[\sqrt{1+z_m^2 A^2}-v(\tret) z_m A\right]+\tret~,\\
X(\tret,z)&=&(z-z_m)\gamma(\tret)\left[v(\tret)\sqrt{1+z_m^2 A^2}- z_m A\right]+x(\tret)~,
\nonumber
\end{eqnarray}
which upon eliminating $\tret$, reduces to\footnote{We thank Eric Pulido for help with this simplification.}
\begin{equation}\label{stringtrajectory}
X(t,z)=\sqrt{A^{-2}+t^2-z^2+z_m^2}~.
\end{equation}
As a (rather trivial) consistency check, notice that the endpoint trajectory $X(t,z_m)$ indeed coincides with (\ref{quarktrajectory}). This solution was also found in \cite{xiao,ppz}, albeit continued all the way down to the AdS boundary $z=0$ and parametrized in terms of the proper acceleration of the corresponding string endpoint, which would be dual to an infinitely massive quark undergoing uniform acceleration. Having here obtained  (\ref{stringtrajectory}) as a special case of (\ref{mikhsolzm}), we are assured that it is the unique retarded embedding that codifies the physics of interest to us.

Combining (\ref{metric}) and (\ref{stringtrajectory}), the induced metric on the worldsheet is found to be
\begin{eqnarray}\label{wsmetric}
h_{tt}&=&-{L^2\over z^2}\left(A^{-2}- z^2+z_m^2 \over A^{-2}+t^2-z^2+z_m^2\right), \\
h_{tz}&=&-{L^2\over z^2}\left(t z \over A^{-2}+t^2-z^2+z_m^2\right), \nonumber\\
h_{zz}&=&{L^2\over z^2}\left(A^{-2}+t^2+z_m^2 \over A^{-2}+t^2-z^2+z_m^2\right).\nonumber
\end{eqnarray}
We see here that $h_{tt}$ vanishes (indicating that the downward-pointing lightcones become horizontal)  at $z=\sqrt{A^{-2}+z_m^2}\equiv z_h$, and so $z_h$ marks the location of a worldsheet horizon. As promised, then, we find that the string embedding dual to a uniformly accelerated quark contains a
worldsheet black hole. The quark/endpoint fluctuations $\delta x(t)$ that we intend to analyze are therefore causally connected (along the worldsheet) only with the $z_m\le z\le z_h$ portion of the string.\footnote{The endpoint can of course also be causally connected to the rest of the string along spacetime trajectories outside the worldsheet, but emission/absorption of the closed string modes that could carry information along such trajectories is suppressed in the large $N_c$ limit.}
The fact that the geometry is static will become manifest in the next section.

It is interesting to note that the solution (\ref{wsmetric}) penetrates into the bulk only up to $z=\sqrt{A^{-2}+t^2+z_m^2}\equiv z_c(t)$. The full string of interest to us must of course extend beyond this radial position, all the way up to $z\to\infty$, in order to be dual to an isolated quark, but has an inflection point at $z=z_c(t)$ (where $\p_z X(t,z_c(t))\to\infty$).
It should be possible to derive the form of the $z>z_c(t)$ portion of the embedding \cite{nolineonthehorizon}, but we will not need it here, because it lies beyond the worldsheet horizon and therefore cannot influence the string endpoint.\footnote{Another possibility is to complete the embedding (\ref{stringtrajectory}) with its reflection across $x=0$, in which case the full string lies at $z\leq z_c(t)$ and is dual to a quark-antiquark pair, with the particles uniformly accelerating back-to-back. Indeed, this is the presentation in which (\ref{stringtrajectory}) was found in \cite{xiao}.}

   \section{Bulk Diffeomorphisms and the Unruh Effect} \label{unruhsec}
   \subsection{Rindler coordinates} \label{rindlersubsec}

  Having determined in the previous section the string embedding dual to a quark that on average follows the uniformly accelerated trajectory (\ref{quarktrajectory}), we would next like to study quantum fluctuations $\delta \vec{x}(t)$ induced by the coupling to the gluonic field. For this purpose, following \cite{rangamani,sonteaney} we must determine the way in which the Hawking radiation emanating from the worldsheet horizon at $z=z_h$ populates the various modes of oscillation of the string, thereby making the endpoint/quark jitter. It will turn out to be easier to address this calculation in coordinates different from those seen in (\ref{wsmetric}), so in this subsection and the next we will present the relevant bulk diffeomorphisms and explain their CFT interpretation. Along the way, we will learn some lessons about the implementation of the Unruh effect in the context of the AdS/CFT correspondence.

  For starters, it is natural to expect our problem to simplify if we switch to a coordinate system adapted to an observer sitting on the quark. To this end, we rewrite the CFT in terms of the Rindler coordinates \cite{unruhreviews,bd}
  \begin{equation}\label{rindlercoordscft}
  t=A^{-1}\exp(A\xp)\sinh(A\tp)~,\qquad x=A^{-1}\exp(A\xp)\cosh(A\tp)~,\qquad \xperp=\xperpp~
  \end{equation}
  ($\xp$ here is the frequently used tortoise longitudinal coordinate, related to Rindler's original choice through $\check{x}^{\prime}\equiv A^{-1}\exp(A\xp)$). As $(\tp,\xp,\xperpp)$ range over the interval $(-\infty,\infty)$, they cover the right Rindler wedge ($x\geq |t|$) of the original $p$-dimensional Minkowski spacetime.  The CFT line element takes the Rindler form
  \begin{equation}\label{rindlermetriccft}
  ds^{2}_{\mathrm{CFT}}=e^{2A\xp}\left(-d\tp^2+d\xp^2\right)+{d\xperpp}^2~.
  \end{equation}
 {}From the point of view of an inertial observer, the worldlines with constant $(\xp,\xperpp)$ describe a family of uniformly accelerated observers with proper acceleration $A\exp(-A\xp)$ and proper time $\tp\exp(A\xp)$,
  so as desired, in the new coordinate system our quark lies at the fixed position $\xp=0$, $\xperpp=0$ and has proper time $\tp$. In accord with the equivalence principle, objects in this frame are immersed in a gravitational field analogous to that of an infinite flat Earth located at the $\xp\to-\infty$ ($\check{x}^{\prime}=0$) plane, and so tend to fall to the left. This effect manifests itself in various ways, and leads in particular to an $\xp$-dependence in the local temperature that we will shortly determine.

  The form of the relation between $t$ and $\tp$ implies that inertial and accelerated observers will disagree in their identification of positive frequency modes, and consequently, in their definition of the CFT vacuum. We will denote the corresponding vacua by  $\ket{\Omega}$ (Minkowski) and $\ket{\Omega^{\prime}}$  (Rindler).
  For the accelerated observers, who follow orbits of the timelike Killing vector $\xi=\p_{\tp}$,
  there is a horizon
   at the edge of the Rindler wedge, $x=|t|$, or equivalently, $\xp\to-\infty$, with surface gravity
   \begin{equation}\label{surfacegravity}
   k_h \equiv -{1 \over 2}\left(\nabla_{\mu} \xi_{\nu} \nabla^{\mu} \xi^{\nu}\right)_{\scriptsize
   \mbox{horizon}}=A~.
   \end{equation}
   As a result, a state that is pure from the inertial perspective will generally be mixed from the point of view of the Rindler observers, because the field degrees of freedom in the Rindler wedge will be entangled with degrees of freedom in the region beyond the horizon, which are traced over. In particular, the Rindler observers will interpret the Minkowski vacuum $\ket{\Omega}$ as a thermal bath  \cite{unruh,unruhreviews,bd} with local temperature
   \begin{equation}\label{localtemp}
   T(\xp)={k_h\over 2\pi \sqrt{-\xi\cdot\xi}}
   =T_{\mathrm{U}}\exp(-A\xp)~,
   \end{equation}
   where
   \begin{equation}\label{unruhtemp}
   T_{\mathrm{U}}={A\over 2\pi}~
   \end{equation}
   denotes the Unruh temperature, which directly gives the temperature of the bath at the (average) location of our quark.

   More specifically, given a set of local operators $\cO_i(x^{\prime\mu})$ evaluated in the right Rindler wedge, the Unruh effect essentially amounts to the statement that\footnote{Strictly speaking, the trace on the right-hand side is not well-defined in the field-theoretic context, so the reinterpretation of the left-hand side as a thermal state must be stated in terms of a KMS condition.}
   \begin{equation}\label{unruhexpec}
   \bra{\Omega}\cO_1(x_1^{\prime\mu})\cdots\cO_n(x_n^{\prime\mu})\ket{\Omega}=
   \tr\left(e^{-H^{\prime}/T_{\mathrm{U}}}\cO_1(x_1^{\prime\mu})\cdots
   \cO_n(x_n^{\prime\mu})\right)
   ~,
   \end{equation}
  where $H^{\prime}\equiv -P_{\tp}=-A(x P_t+t P_x) $ denotes the Rindler Hamiltonian (the generator of translations in $\tp$), and the trace runs over all Rindler states. This equivalence is often discussed explicitly in terms of free scalar fields, but has been proven to hold for an arbitrary interacting field theory on flat space \cite{bisognano}.

  As noted below (\ref{metric}), when AdS is radially foliated in terms of Poincar\'e coordinates, the bulk point $x^{\mu}$ of the leaf at each radial position $z$
  is directly identified with the point $x^{\mu}$ of the CFT, so the bulk transformation dual to (\ref{rindlercoordscft}) is simply
  \begin{eqnarray}\label{rindlercoordsads}
  t&=&A^{-1}\exp(A \xp)\sinh(A\tp)~,\\
  x&=&A^{-1}\exp(A \xp)\cosh(A\tp)~,\nonumber\\
  \xperp&=&\xperpp~,\nonumber\\
  z&=&\zp~,\nonumber
  \end{eqnarray}
  which reexpresses the metric (\ref{metric}) in the form
  \begin{equation}\label{rindlermetricads}
ds^2_{\mathrm{AdS}}={L^2\over \zp^2}\left[e^{2A\xp}\left(-d\tp^2+d\xp^2\right)+{d\xperpp}^2+d\zp^2\right]~.
\end{equation}

{}From (\ref{graltmunu}) it  follows trivially that the spacetime (\ref{rindlermetricads}) describes the CFT state where
\begin{equation}\label{tmunup}
\expec{T^{\,\prime}_{\mu\nu}(\xp)}=0~,
 \end{equation}
 namely, the Minkowski vacuum $\ket{\Omega}$. This assignment might seem to conflict with the statement (\ref{unruhexpec}), but the stress-energy
  of the expected thermal medium does manifest itself in the \emph{difference}
  \begin{equation}\label{tmunumediump}
  \expec{T^{\,\prime}_{\mu\nu}(\xp)}_{\mbox{\scriptsize medium}}\equiv
  \bra{\Omega}T^{\,\prime}_{\mu\nu}(\xp)\ket{\Omega}-\bra{\Omega^{\prime}}T^{\,\prime}_{\mu\nu}(\xp)
\ket{\Omega^{\prime}}
=-\bra{\Omega^{\prime}}T^{\,\prime}_{\mu\nu}(\xp)
\ket{\Omega^{\prime}}~.
\end{equation}
In other words,
the Rindler vacuum $\ket{\Omega^{\prime}}$ is naturally assigned a \emph{negative}  energy density, reflecting the \emph{absence} of the thermal medium \cite{cdeutsch}.
 We will return to this point in the next subsection, where a further bulk transformation will enable us to determine (\ref{tmunumediump}).

In (\ref{rindlermetricads}) we find an acceleration horizon at $\xp\to-\infty$ just like in (\ref{rindlermetriccft}), extending in the radial direction as a reflection of the fact that the CFT horizon equally affects all gluonic modes, from the extreme UV ($z=0$) to the deep IR ($z\to\infty$).
The associated Unruh temperature is again (\ref{unruhtemp}).

In the primed coordinates, the string embedding (\ref{stringtrajectory}) translates into
\begin{equation}\label{stringtrajectoryp}
\Xp(\tp,\zp)={1\over 2A}\ln\left(1-A^2(\zp^2-z_m^2)\right)~,
\end{equation}
which as expected is at rest. The fact that the string is not vertical but bends towards $-\xp$  is a reflection of the fact that a purely radial embedding would not have minimal area: just like in the CFT, trajectories at fixed $\xp$ are non-geodesic, and objects in this frame tend to fall towards the left (aside from being attracted upward, toward larger values of the radial coordinate, which they were already in the unprimed frame).

Since our string lives on the Rindler-AdS spacetime (\ref{rindlermetricads}), we expect it to be exposed to a thermal medium. To see this in more detail, notice that the point  $(t,z)$ on the string embedding (\ref{stringtrajectory}) follows the hyperbolic trajectory $x^2-t^2=A^{-2}-z^2+z_m^2$, which from the CFT perspective corresponds to the $d$-acceleration (defined in (\ref{A})) $A_{\mathrm{CFT}}(z)=1/\sqrt{A^{-2}-z^2+z_m^2}$, and from the bulk perspective corresponds to the $(d+1)$-acceleration
\begin{equation}\label{Aads}
A_{\mathrm{AdS}}(z)\equiv \sqrt{(U^{M}D_{M}U^{N})(U^{P}D_{P}U_{N})}
={1\over L}\sqrt{1+A^2_{\mathrm{CFT}}(z) z^2}~,
\end{equation}
with $D_M$ the bulk covariant derivative and $U^M$ the proper $(d+1)$-velocity. The AdS version of the Unruh effect predicts that an observer undergoing such motion will feel immersed in a thermal medium with local temperature \cite{deser,ppz}
\begin{equation}\label{localtempads}
\mathrm{T}(z)={1\over 2\pi}\sqrt{A^2_{\mathrm{AdS}}(z)-L^{-2}}
={z A_{\mathrm{CFT}}(z)\over 2\pi L}~.
\end{equation}
It is easy to check that, up to the factor of $z/L$ arising from the difference in proper times for the CFT and AdS observers, this is precisely the expected CFT temperature (\ref{localtemp}) at the location $\xp$ assigned to the point $\zp=z$ of the string by the embedding (\ref{stringtrajectoryp}).

The induced worldsheet metric  reads
\begin{eqnarray}\label{wsmetricp}
h_{\tp\tp}&=&-{L^2\over \zp^2}\left(1- A^2(\zp^2-z_m^2)\right), \\
h_{\tp\zp}&=&0, \nonumber\\
h_{\zp\zp}&=&{L^2\over \zp^2}\left(1+ A^2 z_m^2 \over 1- A^2(\zp^2-z_m^2)\right),\nonumber
\end{eqnarray}
corresponding to a manifestly static black hole with horizon at $\zp=z_h$,
and associated Unruh/Hawking temperature again given by (\ref{unruhtemp}).

   \subsection{Removing the Rindler horizon via a change of conformal frame} \label{conformalsubsec}

   The presence of a Rindler horizon in the CFT arises from the $\exp(2A\xp)$ factor in the line element (\ref{rindlermetriccft}), so the fact that we are dealing with a conformal theory presents us with an interesting possibility: we can remove this factor from the $(\tp,\xp)$ portion of the metric via a change of conformal frame.
    To be more precise, among the infinite number of different conformal frames available to an observer sitting on the quark, we can choose the one related to (\ref{rindlermetriccft}) via the specific Weyl transformation $ds^{2}_{\mathrm{CFT}}\to \exp(-2A\xp)ds^{2}_{\mathrm{CFT}}$, to be left with
   \begin{equation}\label{openeinsteinmetriccft}
  ds^{2}_{\mathrm{CFT}}=-d\tpp^2+d\xpp^2+e^{-2A\xpp}{d\xperppp}^2~.
  \end{equation}
  By definition, Weyl transformations locally rescale the metric while leaving the coordinates untouched;\footnote{For maximal clarity, we note that, in the GR literature, these mappings are  directly called conformal transformations, but it is frequent in the CFT and string theory literature to reserve the latter denomination for mappings that leave the metric untouched
  while pushing the points of the manifold around in a way that preserves angles. In either presentation, these transformations generally induce position-dependent rescalings of proper lengths, and are therefore distinct from conformal diffeomorphisms (or conformal isometries), which transform both the metric and the coordinates leaving proper lengths invariant. A conformal transformation, in the second sense of the phrase, can always be pictured as a conformal diffeomorphism composed with a Weyl transformation chosen to bring the metric back to its original form. It follows then that, in any theory invariant under diffeomorphisms, Weyl invariance implies conformal invariance, but the converse holds only in 2 dimensions, because for $d>2$ conformal transformations constitute only a finite-dimensional group (e.g., $SO(d,2)$ for conformally flat metrics).}
  nevertheless, we have relabeled the coordinates with double primes instead of primes because this will shortly prove convenient. The line element (\ref{openeinsteinmetriccft}) is that of the open Einstein universe, $\bR\times\mathbf{H}^{d-1}$. Our parametrization of the hyperbolic space is related to the standard Poincar\'e (or horospheric) coordinates through $\check{x}^{\prime\prime}\equiv A^{-1}\exp(A\xpp)$, and to the perhaps more familiar ($k=-1$) static Robertson-Walker form of the line element through \cite{cdowker}
  \begin{eqnarray}\label{robertsonwalker}
  \check{x}^{\prime\prime}&=&\frac{A^{-1}} {\cosh{\chi^{\prime\prime}}-\sinh{\chi^{\prime\prime}}\cos\theta^{\prime\prime}_1}~,\nonumber\\
  x^{\prime\prime}_2&=&\frac{A^{-1}\sinh{\chi^{\prime\prime}}\sin{\theta^{\prime\prime}_1}
  \cos{\theta^{\prime\prime}_2}}
  {\cosh{\chi^{\prime\prime}}-\sinh{\chi^{\prime\prime}}\cos\theta^{\prime\prime}_1}~,\\
  &\vdots& \nonumber\\
  x^{\prime\prime}_{d-1}&=&\frac{A^{-1}\sinh{\chi^{\prime\prime}}
  \sin{\theta^{\prime\prime}_1}\cdots\sin{\theta^{\prime\prime}_{d-2}}}
 {\cosh{\chi^{\prime\prime}}-\sinh{\chi^{\prime\prime}}\cos\theta^{\prime\prime}_1}~.\nonumber
 \end{eqnarray}

  Naively, the possibility of removing the horizon might make it seem like the interpretation of the Minkowski vacuum as a thermal state depends on the choice of conformal frame, in which case the Unruh effect would somehow not be fully present in a CFT. This issue was explored in the free field context a couple of decades ago \cite{cdowker,bd}. We will now show that new light is shed on this question for the case of strongly-coupled CFTs when we examine it on the AdS side of the correspondence.

  As was first explained in detail in \cite{theisen}, arbitrary Weyl transformations in the CFT are dual to bulk diffeomorphisms which preserve the Fefferman-Graham form (\ref{feffermangraham}) of the bulk metric. Concretely, $g_{\mu\nu}(x)\to \exp(2\omega(x))g_{\mu\nu}(x)$ corresponds to a bulk transformation involving $z\to \exp(\omega(x))z$ and a compensating change in $x^{\mu}$ that prevents the appearance of a $z$-$\mu$ cross-term in the bulk metric. (So, taking this family of bulk diffeomorphisms into account, a given bulk metric is understood to induce not a specific boundary metric, but a specific boundary conformal structure \cite{w}.) In particular, the Weyl transformation leading from (\ref{rindlermetriccft}) to (\ref{openeinsteinmetriccft}) is dual to a bulk diffeomorphism that changes the radial foliation according to\footnote{For later convenience, we also include here a rigid rescaling of the transverse directions.}
 \begin{eqnarray}\label{openeinsteincoordsads}
  \tp&=&\tpp~,\\
  \xp&=&\xpp+A^{-1}\ln\left(\sqrt{1+A^2 z_m^2}\sqrt{1-A^2\zpp^2}\right)~,\nonumber\\
  \xperpp&=&\sqrt{1+A^2 z_m^2}\,\xperppp~,\nonumber\\
  \zp&=&\sqrt{1+A^2 z_m^2}\,\zpp\exp(A\xpp)~.\nonumber
  \end{eqnarray}
This converts the metric (\ref{rindlermetricads}) into
\begin{equation}\label{openeinsteinmetricads}
ds^2_{\mathrm{AdS}}={L^2\over \zpp^2}\left[-\left(1- A^2\zpp^2\right)d\tpp^2+d\xpp^2
+\,e^{-2A\xpp}{d\xperppp}^2+\frac{d\zpp^2}{ 1- A^2\zpp^2}\right]~,
\end{equation}
which indeed gives rise to (\ref{openeinsteinmetriccft}) at the boundary.\footnote{And can be brought back to Fefferman-Graham form through the trivial bulk diffeomorphism $\zpp=\zt/(1+A^2\zt^2/4)$.}

It is clear from this line element that the acceleration horizon is still present in the bulk, and still has an associated Unruh temperature (\ref{unruhtemp}), but now lies on the fixed radial plane $\zpp=A^{-1}=2\pi/T_{\mathrm{U}}$, which explains why it is no longer visible as a horizon from the CFT perspective. In other words, in this new conformal frame, the thermal character of the CFT state arises not from entanglement with degrees of freedom that lie beyond a spacetime horizon, but from the direct identification of the specific energy scale $T_{\mathrm{U}}$ as the temperature of the CFT.

For the case $d=2$ (where we are dealing with AdS$_3$), and under an appropriate periodic identification of the $\xpp$ coordinate, the metric (\ref{openeinsteinmetricads}) is that of the BTZ black hole \cite{btz,bhtz}. In our case, however, $\xpp$ is non-compact, and so what we have, for all values of $d$, is naturally not a black hole but just a portion of AdS$_{d+1}$.

Using (\ref{graltmunu}), we can deduce from (\ref{openeinsteinmetricads}) the expectation value of the stress-energy tensor of the CFT in the open Einstein universe (\ref{openeinsteinmetriccft}),
\begin{equation}\label{tmunupp}
\expec{T^{\prime\prime}_{\mu}{}^{\,\nu}} = \left\{\!\!\!\!\!\!\!\!\!\!
\begin{minipage}{8cm}\vspace{-0.4cm}
\begin{eqnarray}
&{}&{} \quad 0 \quad\qquad\qquad\qquad\qquad\qquad\qquad d\;\mbox{odd}~,
\nonumber\\
&{}&{}{{  L A^2 \over 16\pi G^{(3)}_{\mbox{\scriptsize N}}}\,\mbox{diag}(-1,1)\qquad\qquad\qquad d=2}~,
\nonumber\\
&{}&{}{{ L^3 A^4 \over 64\pi G^{(5)}_{\mbox{\scriptsize N}}}\,\mbox{diag}(-3,1,1,1)\qquad\qquad d=4}~,
\nonumber\\
&{}&{}{{ L^5 A^6 \over 128\pi G^{(7)}_{\mbox{\scriptsize N}}}\,\mbox{diag}(-5,1,1,1,1,1)\,\;\;\quad d=6}~.
\nonumber
\end{eqnarray}
\end{minipage}
\right.
\end{equation}
This is in complete agreement with the results of \cite{roberto}, where these stress tensors were obtained in the context of a more general investigation of hyperbolic black holes in AdS/CFT (whose conclusions will be of further use to us below).\footnote{The case $d=4$ was also considered very recently in \cite{hirayama}, which appeared while the present paper was in preparation, but the result of that work differs from ours by an overall factor of $\exp(-2A\xpp)/2$.}

We notice from (\ref{tmunupp}) that, in even dimensions, the expectation value is nonvanishing in spite of the fact that we are in the state that is conformally related to the Minkowski vacuum. The reason is of course well-understood: our CFT is classically Weyl invariant, but at the quantum level this symmetry is anomalous precisely for even $d$ \cite{duff} (for reviews see, e.g., \cite{bd,duffreview}). This is evidenced by the fact that, when the theory is defined on a curved background with metric $g_{\mu\nu}$, the trace of the energy-momentum tensor is generically non-zero,
\begin{equation}\label{weylanomaly}
\expec{T_{\mu}{}^{\mu}}_{g} = \left\{\!\!\!\!\!\!\!\!\!\!
\begin{minipage}{9cm}\vspace{-0.4cm}
\begin{eqnarray}
&{}&{}{{ c \over 24\pi }R \;\;\quad\qquad\qquad\qquad\qquad\qquad\qquad\qquad\qquad d=2}~,
\nonumber\\
&{}&{}{ 1 \over 16\pi^2 }\left[
\alpha\left(R_{\mu\nu\lambda\rho}R^{\mu\nu\lambda\rho}-2R_{\mu\nu}R^{\mu\nu}+{1\over 3}R^2\right)
\right. \nonumber\\
&{}&{} \left. \quad\qquad +\beta\bigg(R_{\mu\nu\lambda\rho}R^{\mu\nu\lambda\rho}-4R_{\mu\nu}R^{\mu\nu}+R^2\bigg)
\right]\quad d=4 ~,
\nonumber
\end{eqnarray}
\end{minipage}
\right.
\end{equation}
(and similarly for $d=6$),
where $c,\alpha,\beta$ are numerical constants that depend on the field content.

One of the most impressive pieces of evidence supporting the AdS/CFT correspondence is the fact that the Weyl anomaly can be reproduced from the dual classical gravity setup, as was first demonstrated in \cite{hs} (following a suggestion of \cite{w}). Indeed, the energy-momentum tensor (\ref{graltmunu}), derived by functional differentiation of the appropriately renormalized gravity action (i.e., Einstein-Hilbert plus counterterms chosen to eliminate the IR divergences), has a trace that agrees with (\ref{weylanomaly}), with
\begin{equation}\label{weylcoeffs}
c={3 L \over 2 G^{(3)}_{\mbox{\scriptsize N}}}~,\quad
\alpha= {\pi L^3 \over 8 G^{(5)}_{\mbox{\scriptsize N}}}=-\beta~.
\end{equation}
These coefficients can be shown to coincide with those expected at weak coupling (and in the large $N_c$ limit) for the field content of the CFT, in every case where the latter is known (e.g., for $\cN=4$ SYM, $\alpha=-\beta=N_c^2/4$) \cite{hs}.
On the gravity side, the Weyl anomaly translates into the statement that,
generically, the diffeomorphisms dual to Weyl transformations of the CFT \cite{theisen} are not true symmetries of the theory.

It is easy to check that, for the Rindler metric (\ref{rindlermetriccft}) and open Einstein metric (\ref{openeinsteinmetriccft}), the geometric expression for the Weyl anomaly vanishes (the first case is flat, and hence trivial). This is consistent with the fact that the corresponding stress tensors (\ref{tmunup}) and (\ref{tmunupp}) are traceless, and implies that, starting with either of these backgrounds, \emph{infinitesimal} Weyl transformations are true symmetries. But in going from the primed to the doubly primed frame we have made a \emph{finite} Weyl transformation, which is \emph{not} a true symmetry because the anomaly happens to be non-zero for all metrics `in between'. This is the origin of the anomalous shift in the energy-momentum tensor.

In more detail, for a theory that is classically Weyl-invariant, the transformation $\bar{g}\to g\equiv\exp(2\omega)\bar{g}$, with $\bar{g}$ a flat metric,  induces a change in the energy-momentum tensor that follows directly from integration of the Weyl anomaly. E.g., for $d=4$ \cite{cc,bd},
\begin{equation}\label{tmunutransform}
\expec{T_{\mu}{}^{\nu}}_{g}= \sqrt{\,\bar{g}\,\over \,g\,}\expec{T_{\mu}{}^{\nu}}_{\bar{g}}
+{1\over 16\pi^2}\left[{1\over 9}\alpha H^{(1)}_{\mu}{}^{\,\nu} + 2\beta H^{(3)}_{\mu}{}^{\,\nu}\right]~,
\end{equation}
where
\begin{eqnarray}\label{H}
H^{(1)}_{\mu\nu}&\equiv &-2\del\!_{\mu}\del\!_{\nu}R -2g_{\mu\nu}\del^2 R +{1\over 2}g_{\mu\nu}R^2-2RR_{\mu\nu}~,\nonumber\\
H^{(3)}_{\mu\nu}&\equiv &-R_{\mu}{}^{\rho}R_{\rho\nu}+{2\over 3}R R_{\mu\nu}+{1\over 2}R_{\lambda\rho}R^{\lambda\rho}g_{\mu\nu}-{1\over 4}R^2 g_{\mu\nu}~.
\end{eqnarray}
Since AdS/CFT correctly reproduces the Weyl anomaly, the energy-momentum tensor (\ref{graltmunu}) should automatically transform in this manner. This was verified in \cite{dhss} for infinitesimal Weyl transformations about an arbitrary metric. In our setting, given the fact that $\expec{T^{\,\prime}_{\mu\nu}}=0$ in the Minkowski vacuum, and the form of the metrics (\ref{rindlermetriccft}) and (\ref{openeinsteinmetriccft}), it is straightforward to verify that the doubly primed tensor (\ref{tmunupp}) indeed has the form expected from the finite Weyl transformation, and in particular satisfies (\ref{tmunutransform}) when $d=4$.

The shift in the energy-momentum tensor induced by the Weyl anomaly is purely geometric, and therefore independent of the state in which the expectation value is computed. In particular, (\ref{tmunutransform})
holds equally for the Minkowski vacuum $\ket{\Omega}$ and the Rindler vacuum $\ket{\Omega^{\prime}}$. This implies that the stress tensor of the Unruh thermal medium, defined in (\ref{tmunumediump}) as the \emph{difference} between the Minkowski and Rindler expectation values, is related to the corresponding difference in the open Einstein universe simply through
\begin{equation}\label{tmunumediumpfrompp}
  \expec{T^{\,\prime}_{\mu}{}^{\,\nu}}_{\mbox{\scriptsize medium}}=
  \sqrt{\,g^{\prime\prime}\,\over\,g^{\prime}\,}\bigg(
  \bra{\Omega}T^{\prime\prime}_{\mu\nu}\ket{\Omega}-\bra{\Omega^{\prime}}T^{\prime\prime}_{\mu\nu}
\ket{\Omega^{\prime}}\bigg)~.
\end{equation}

The conformal Minkowski vev in the right-hand side of the preceding expression is already available to us in (\ref{tmunupp}). Direct field-theoretic analysis shows that the conformal Rindler vev vanishes for $d=4$ \cite{bunch}.
The corresponding result for arbitrary $d$ was deduced in \cite{roberto} via AdS/CFT. That paper considered an infinite family of hyperbolic black holes in AdS$_{d+1}$ parametrized by their temperature. The translation of these solutions to our language involves the usual inversion of the radial coordinate, $r\to z=L^2/r$, followed by the bulk diffeomorphism $x^{\mu}/R\to Ax^{\mu}$, $z/R\to Az$ (which is dual to a conformal transformation in the CFT), to be left with
 \begin{eqnarray}\label{robertometric}
 ds^2_{\mathrm{AdS-BH}}&=&{L^2\over \zpp^2}\left[-\left(1- A^2\zpp^2-{\mu\over L^{d-2}}A^d\zpp^d\right)d\tpp^2+d\xpp^2+\,e^{-2A\xpp}{d\xperppp}^2 \right.
 \nonumber\\
{}&{}&\quad\qquad\left.+\frac{d\zpp^2}{ 1- A^2\zpp^2-{\mu\over L^{d-2}}A^d\zpp^d}\right]~,
 \end{eqnarray}
 with $\mu$ a mass-density parameter that controls the black hole temperature $T_{\mbox{\scriptsize BH}}$. The case $\mu=0$ clearly coincides with our pure AdS metric (\ref{openeinsteinmetricads}), and was indeed identified in \cite{roberto} as dual to the conformal Minkowski vacuum of the CFT on the open Einstein universe. The zero temperature, extremal, solution has $\mu/L^{d-2}=-2(d-2)^{d/2-1}/d^{d/2}$, and was argued in \cite{roberto} to be dual to the conformal Rindler vacuum.\footnote{More generally, the geometry (\ref{robertometric}) is dual to the state of the open Einstein CFT conformal to a thermal ensemble of the Rindler observer at temperature $T_{\mbox{\scriptsize BH}}$.} The corresponding stress tensor is
 \begin{equation}\label{tmunurindlerpp}
 \bra{\Omega^{\prime}}T^{\prime\prime}_{\mu\nu}
\ket{\Omega^{\prime}}=
{ L^{d-1} A^d \over 16\pi G^{(d+1)}_{\mbox{\scriptsize N}}}
\left({2\eps^0_{(d)}\over d-1}-{2(d-2)^{d/2-1}\over d^{d/2}}\right)\,\mbox{diag}(d-1,1,\ldots,1)~,
 \end{equation}
 with $\eps^0_{(d)}=0$ for odd $d$, and $\eps^0_{(d)}=(d-1)!!^2/d!$ for even $d$.

 Using (\ref{tmunupp}) and (\ref{tmunurindlerpp}) in (\ref{tmunumediumpfrompp}) we can finally deduce the stress-energy of the Unruh thermal medium detected in the Minkowski vacuum by the Rindler CFT observer,\footnote{This result was also implicit in \cite{roberto}, but the Unruh effect was not a central concern in that work.}
 \begin{equation}\label{tmunumediumresult}
 \expec{T^{\,\prime}_{\mu}{}^{\,\nu}}_{\mbox{\scriptsize medium}} =
{(d-2)^{d/2-1} L^{d-1} A^d \over 8\pi d^{d/2} G^{(d+1)}_{\mbox{\scriptsize N}}}
e^{-d A \xpp}\,\mbox{diag}(d-1,1,\ldots,1)~.
 \end{equation}
We notice here the expected divergence ($\propto \check{x}^{\prime\prime}{}^{-d}$) of the local energy density as the Rindler horizon is approached \cite{cdeutsch}.

Now that we
understand the precise meaning of our coordinate transformation, let us shift our focus back to
the behavior of the quark/string.
In the doubly primed coordinates, the string embedding (\ref{stringtrajectoryp}) is simply
\begin{equation}\label{stringtrajectorypp}
\Xpp(\tpp,\zpp)=0~,
\end{equation}
which is not only static but also vertical. This reflects the fact that in the new conformal frame there is no longer a gravitational potential pulling objects to the left.
Notice that the spacetime geometry (\ref{openeinsteinmetricads}) at the location of the string, which is all that is relevant for the small string perturbations that we will compute in the next section, makes no distinction between the directions $\xpp$ and $\xperppp$, so the dual quark is immersed in a thermal medium that (to first order) is  isotropic. The string endpoint is at $\zpp=z_m/\sqrt{1+A^2 z_m^2}\equiv z_m^{\,\prime\prime}$. As explained in \cite{martinfsq,damping}, in the CFT the length $z_m^{\,\prime\prime}$ gives the characteristic size of the `gluonic cloud' surrounding the quark, or in other words, the quark Compton wavelength.

 The induced worldsheet metric now directly coincides with the $(\tpp,\zpp)$ block of the spacetime metric,
\begin{eqnarray}\label{wsmetricpp}
h_{\tpp\!\tpp}&=&-{L^2\over \zpp^2}\left(1- A^2\zpp^2\right), \\
h_{\tpp\!\zpp}&=&0, \nonumber\\
h_{\zpp\!\zpp}&=&{L^2\over \zpp^2}\left(1- A^2\zpp^2\right)^{-1},\nonumber
\end{eqnarray}
and so displays the same horizon at $\zpp=A^{-1}\equiv z^{\prime\prime}_h$.
Notice, however, that, because the worldsheet is only 2-dimensional, the line $\zpp=A^{-1}$ (presented either in the unprimed, primed or doubly primed coordinates) is an event horizon, and the region behind it is a true black hole.
 The small fluctuations  $\delta\vec{X}^{\,\prime\prime}(\tpp,\zpp)$ that are of interest to us are free massless scalar fields that propagate on this geometry, and, just like in \cite{rangamani,sonteaney}, get excited by the Hawking radiation emanating from the horizon.

   Combining (\ref{rindlercoordsads}) and (\ref{openeinsteincoordsads}), the relation between the unprimed and doubly primed coordinates is seen to be
     \begin{eqnarray}\label{btztransform}
  t&=&A^{-1}\sqrt{1+A^2 z_m^2}\sqrt{1- A^2\zpp^2}\exp(A \xpp)\sinh(A\tpp)~,\\
  x&=&A^{-1}\sqrt{1+A^2 z_m^2}\sqrt{1- A^2\zpp^2}\exp(A \xpp)\cosh(A\tpp)~,\nonumber\\
    \xperp&=&\sqrt{1+A^2 z_m^2}\,\xperppp~,\nonumber\\
  z&=&\sqrt{1+A^2 z_m^2}\,\zpp\exp(A\xpp)~.\nonumber
  \end{eqnarray}
  This transformation was written down already in \cite{bhtz},
  and considered in the AdS/CFT context in \cite{xiao} (as well as in the very recent work \cite{hirayama}, which appeared while the present paper was in preparation), but its precise CFT interpretation had not been previously elucidated.

\section{Quantum Fluctuations of the Accelerating Quark} \label{fluctuationssec}
   \subsection{Langevin equation in the Unruh medium} \label{langevinunruhsubsec}

We are now ready to determine the form of the quantum fluctuations of the string endpoint, by examining how small perturbations of the string embedding get excited by Hawking radiation emanating from the worldsheet black hole. Clearly it will be easier to perform this calculation in either the primed or doubly primed coordinate systems defined in the previous section, where the black hole geometry is manifestly static. {}From the analysis of the previous section we know that the doubly primed setup should be directly dual to a static quark in a
 thermal medium that to zeroth order looks isotropic at the location of the quark, so it is natural to expect the computation in this system to run in complete parallel with the analyses of thermal fluctuations in \cite{rangamani,sonteaney}.

 As a matter of fact, we can notice that, for the case $d=2$, the string embedding (\ref{stringtrajectorypp}) and the spacetime metric (\ref{openeinsteinmetricads}) at the location of the string coincide \emph{exactly} with the corresponding metric and embedding considered in Section 2.2 of \cite{rangamani}, under the identifications
 \begin{equation} \label{rangamani}
 \tpp=t_{\mathrm{there}}~,\quad \xpp=x_{\mathrm{there}}~,\quad \zpp={\ell^2\over r}~,
 \quad L=\ell~, \quad A={r_H\over\ell^2}~, \quad z_m^{\,\prime\prime}={\ell^2\over r_c}~.
 \end{equation}
 The embedding of the flavor branes is different: whereas in \cite{rangamani} they cover the region of AdS between the boundary and the $\zpp= z_m^{\,\prime\prime}$ plane, in our case they extend up to
 \begin{equation}\label{flavorbranes}
 \zpp=z_m^{\,\prime\prime}e^{-A\xpp}~,
 \end{equation}
which is the image of the unprimed locus $z=z_m$. This implies that at the average position of its endpoint, $\xpp=0$, our string should satisfy a Neumann boundary condition along the line $\zpp=z_m^{\,\prime\prime}(1-A\xpp)$, instead of simply along $\zpp$. This distinction, however, is itself of first order in the perturbation about the string embedding (\ref{stringtrajectorypp}), and consequently negligible at our level of approximation. For $d=2$, we can therefore carry over to our setting the worldsheet perturbation analysis of \cite{rangamani} and the resulting Langevin equation describing the evolution of thermal fluctuations of the quark along the single spatial direction $\xpp$.

For $d>2$, our doubly primed metric (\ref{openeinsteinmetricads}) differs from the corresponding metric in \cite{rangamani}, because the latter involves exponents that depend on $d$. But the only novelty in our calculation is the presence of the transverse fluctuations $\delta \Xperppp(\tpp,\zpp)$, which, given the isotropy of the spacetime metric (\ref{openeinsteinmetricads}) at the location of the string, must evolve in exactly the same way as the longitudinal fluctuations $\delta \Xpp(\tpp,\zpp)$, whose behavior is already known to us via the contact with the $d=2$ case of \cite{rangamani}.

For arbitrary dimension, then, we find that as a result of the thermal medium present in the doubly primed system, the position of the quark fluctuates in such a way that both longitudinal and transverse perturbations obey the generalized non-relativistic  Langevin equation
\begin{equation} \label{langevinpp}
m^{\prime\prime} {d^2\delta x_i^{\prime\prime}\over d\tpp^2}(\tpp)
+\int_{-\infty}^{\tpp}\! ds^{\prime\prime}\, \eta^{\prime\prime}\!(\tpp\!-\! s^{\prime\prime}){d\delta x_i^{\prime\prime}\over d s^{\prime\prime}}(s^{\prime\prime})
=f_i^{\,\prime\prime}\!(\tpp)
~,
\end{equation}
where $m^{\prime\prime}\equiv \sqrt{\lambda}/2\pi z^{\,\prime\prime}_m=\sqrt{1+A^2 z_m^2}\,m$
and $f_i^{\,\prime\prime}$ is a random force with statistical averages
\begin{equation}\label{fpp}
\expec{f_i^{\,\prime\prime}(\tpp)}=0~, \qquad
\expec{f_i^{\,\prime\prime}(\tpp)f_j^{\,\prime\prime}(s^{\prime\prime})}
=\delta_{i j}\kappa^{\prime\prime}\!(\tpp\!-\! s^{\prime\prime})~.
\end{equation}
The friction kernel $\eta^{\prime\prime}(\tpp)$ and the stochastic force correlator $\kappa^{\prime\prime}\!(\tpp)$ (respectively denoted $m\gamma(t)$ and $\kappa(t)$ in \cite{rangamani})
can be specified in terms of their Fourier transforms\footnote{Notice that, by causality, $\eta^{\prime\prime}\!(\tpp)$ is understood to vanish for negative argument.}
\begin{eqnarray}\label{etakappathermal}
\eta^{\prime\prime}(\omega)&\equiv&\int_0^{\infty}\! d\tpp\,\eta^{\prime\prime}\!(\tpp)e^{i\omega\tpp}
=im^{\prime\prime}\omega+{\sqrt{\lambda}\over 2\pi}\left(\frac{4\pi^2 T_{\mathrm{U}}^2-i{2\pi\over\sqrt{\lambda}}m^{\prime\prime}\omega}
{1-i{\sqrt{\lambda}\omega\over 2\pi m^{\prime\prime}}}\right)~,\\
 \kappa^{\prime\prime}(\omega)&\equiv&\int_{-\infty}^{\infty}\! d\tpp\,\kappa^{\prime\prime}\!(\tpp)e^{i\omega\tpp}
 ={\sqrt{\lambda}\over\pi}\left(\frac{4\pi^2 T_{\mathrm{U}}^2+\omega^2}
 {1+{\lambda\omega^2\over 4\pi^2m^{\prime\prime 2}}}\right)
 \frac{|\omega|}{\exp({|\omega|\over T_{\mathrm{U}}})-1} ~. \nonumber
\end{eqnarray}
These two functions are connected through the fluctuation-dissipation relation \cite{rangamani,sonteaney} $2\mathrm{Re}\,{\eta^{\prime\prime}(\omega)}=[(\exp({|\omega|/T_{\mathrm{U}}})-1)/|\omega|]
\kappa^{\prime\prime}(\omega)$.
If we study the fluctuations over time scales much larger than the quark Compton wavelength, $\Delta\tpp\gg z^{\,\prime\prime}_m$,
we can approximate (\ref{etakappathermal}) by its low-frequency form ($\omega\ll 2\pi m^{\prime\prime}/\sqrt{\lambda}$), and then (\ref{langevinpp})  reduces to the standard local Langevin equation with white noise,
\begin{equation} \label{locallangevinpp}
m^{\prime\prime}_{\mathrm{th}} {d^2\delta x_i^{\prime\prime}\over d\tpp^2}(\tpp)
+\eta^{\prime\prime}_0{d\delta x_i^{\prime\prime}\over d\tpp}(\tpp)
=f_i^{\,\prime\prime}\!(\tpp)
~,\quad
\expec{f_i^{\,\prime\prime}(\tpp)f_j^{\,\prime\prime}(s^{\prime\prime})}
=\delta_{i j}\kappa^{\prime\prime}_0\delta(\tpp\!-\! s^{\prime\prime})~,
\end{equation}
where $\eta^{\prime\prime}_0\equiv \eta^{\prime\prime}(\omega=0)=2\pi\sqrt{\lambda} T_{\mathrm{U}}^2$, $\kappa^{\prime\prime}_0\equiv \kappa^{\prime\prime}(\omega=0)=4\pi\sqrt{\lambda} T_{\mathrm{U}}^3$
and\footnote{This last definition is a bit of a puzzle to us. We
 expected the linear term in the low frequency approximation to $\eta(\omega)$ to
 contribute to the mass term in (\ref{locallangevinpp}) with the known thermal correction, as was
 verified in \cite{sonteaney} for the case $d=4$, but that would have led to $$m^{\prime\prime}_{\mathrm{th}}=m^{\prime\prime}(1-z^{\prime\prime}_m/z^{\prime\prime}_h)
 =m^{\prime\prime}(1-\sqrt{\lambda}T_{\mathrm{U}}/m^{\prime\prime}).$$
 There is presumably some slight error either in the results of \cite{rangamani} or in our interpretation of them,
 but, regrettably, we have not been able to find it. We should also note that the questionable quadratic temperature-dependence seen in the thermal mass shift deduced by us has a common origin with the quadratic dependence
 in the friction coefficient $\eta^{\prime\prime}_0$, which is definitely correct.}
 $m^{\prime\prime}_{\mathrm{th}}\equiv m^{\prime\prime}(1-\lambda T_{\mathrm{U}}^2/m^{\prime\prime\,2})$.
The force correlation strength $\kappa^{\prime\prime}_0$ obtained here coincides with the result derived previously in \cite{xiao}, in the context of a momentum broadening computation.\footnote{We are grateful to Bo-Wen Xiao for pointing out that the numerical disagreement reported in v1 of this paper on the arXiv is actually nonexistent.}

\subsection{Langevin equation for an inertial observer} \label{langevininertialsubsec}

Having understood the way in which the thermal (Unruh) medium in the doubly primed frame makes our quark fluctuate, we can now transform back to the unprimed, inertial frame, to obtain the radiation-induced quantum fluctuations which are our main interest. For this purpose we need to relate the corresponding  fluctuating string embeddings
$$
X_i(t,z)=\sqrt{A^{-2}+t^2+z_m^2-z^2}\,\delta_{i1}+\delta X_i(t,z)
\quad\longleftrightarrow\quad
X^{\prime\prime}_i(\tpp,\zpp)=0+\delta X^{\prime\prime}_i(\tpp,\zpp),
$$
a task which requires two separate steps. First, using (\ref{btztransform}) evaluated at the location of the string, $\xpp=0$, we recognize that the coordinates on the worldsheet transform according to
\begin{equation}\label{wstransform}
\tpp=A^{-1}\mathrm{arcsinh}\!\left({At\over\sqrt{1-A^2(z^2-z_m^2)}}\right)~,\qquad
\zpp={z\over\sqrt{1+A^2 z_m^2}}~.
\end{equation}
When evaluated at the string endpoint, the first relation tells us that $\tpp$ agrees with the quark proper time (\ref{tau}), as it should. So the time derivatives in (\ref{langevinpp}) or (\ref{locallangevinpp}) are connected to their inertial counterparts via $d/d\tpp=\sqrt{1+A^2 t^2}d/dt$. Next, at any given point $(t,z)\leftrightarrow(\tpp,\zpp)$ on the string worldsheet, we can perturb (\ref{btztransform}) to conclude that
\begin{equation}\label{flucttransform}
\delta X =\frac{(1+A^2 z_m^2)\sqrt{1+A^{2}(t^2+z_m^2-z^2)}}{1-A^2(z^2-z_m^2)}\,\delta X^{\prime\prime}~,\quad
\delta \vec{X}_{\perp}= \sqrt{1+A^2 z_m^2}\,\delta \vec{X}^{\prime\prime}_{\perp}~.
\end{equation}
Evaluating at the string endpoint, this tells us how to relate the quark fluctuations $\delta x_i$ to $\delta x^{\prime\prime}_i$.

The only other element needed to complete the translation of  (\ref{langevinpp}) into the unprimed frame is  the transformation rule for the forces, which can be obtained as follows. As we have done all along, let $\vec{v}$ and $\gamma$ denote the velocity and Lorentz factor of the quark undergoing the hyperbolic motion (\ref{quarktrajectory}), and $\vec{F}$ the corresponding external $(d-1)$-force, determined by (\ref{A}). Our quark is subjected to this force in combination with the radiation-induced stochastic force $\vec{f}$ that we are now trying to determine, and  as a result acquires a perturbed velocity $\vec{v}_{\mbox{\scriptsize{tot}}}\equiv \vec{v}+\delta\vec{v}$ and corresponding Lorentz factor $\gamma_{\mbox{\scriptsize{tot}}}$. We will denote the associated $d$-force by
$\cF^{\mbox{\scriptsize{CFT}}}_{\mu}=\gamma_{\mbox{\scriptsize{tot}}}
(-\vec{F}_{\mbox{\scriptsize{tot}}}\cdot\vec{v}_{\mbox{\scriptsize{tot}}},
\vec{F}_{\mbox{\scriptsize{tot}}})$. Knowing that
\begin{equation}\label{calFtransformCFTtoAdS}
\cF^{\mbox{\scriptsize{AdS}}}_{M}=
\left(\frac{d\tau_{\mbox{\scriptsize{CFT}}}}{d\tau_{\mbox{\scriptsize{AdS}}}}
\cF^{\mbox{\scriptsize{CFT}}}_{\mu},0\right)
\end{equation}
(where $d\tau_{\mbox{\scriptsize{CFT}}}$ and $d\tau_{\mbox{\scriptsize{AdS}}}=(L/z)d\tau_{\mbox{\scriptsize{CFT}}}$ respectively denote the CFT and AdS proper times) transforms as a $(d+1)$-vector, we can deduce that
\begin{equation}\label{calFtransformCFTtoCFT}
\cF^{\prime\prime\mbox{\scriptsize{CFT}}}_{\mu}=
\frac{\p x^{\nu}}{\p x^{\prime\prime\mu}}\cF^{\mbox{\scriptsize{CFT}}}_{\nu}~.
\end{equation}
Taylor-expanding this relation to first order in the perturbation, we find that
\begin{equation}\label{ftransform}
f^{\prime\prime}=f~,\qquad \vec{f}^{\,\prime\prime}_{\perp}=\sqrt{1+A^2 z_m^2}\sqrt{1+A^2 t^2}\,\vec{f}_{\perp}~.
\end{equation}

Putting all of this together, we finally conclude that, in the original inertial frame, the radiation emitted by the quark induces quantum longitudinal and transverse fluctuations respectively obeying the differential equations
\begin{equation} \label{langevinlong}
m {d^2\delta x\over d t^2}(t)
+\int_{-\infty}^{t}\! ds\, \left(\eta(t,s){d\delta x \over d s }(s)
-\zeta(t,s)\delta x(s) \right)
=\frac{\sqrt{1+{\lambda A^2 \over 4\pi^2 m^2}}f(t)}{\sqrt{1+A^2 t^2}}~
\end{equation}
and
\begin{equation} \label{langevinperp}
m {d^2\delta\xperp\over d t^2}(t)
+\int_{-\infty}^{t}\! ds\, \eta_{\perp}(t,s){d\delta\xperp\over d s }(s)
=\frac{\sqrt{1+{\lambda A^2 \over 4\pi^2 m^2}}\vec{f}_{\perp}(t)}{\sqrt{1+A^2 t^2}}~,
\end{equation}
with
\begin{equation}\label{f}
\expec{f_i(t)}=0~, \qquad
\expec{f_i(t)f_j(s)}
=\delta_{i j}\kappa_i(t,s) ,
\end{equation}
where we have defined
\begin{eqnarray}\label{etakappa}
\eta(t,s)&\equiv& \frac{\eta^{\prime\prime}\!\left(A^{-1}\mathrm{arcsinh}(At)-A^{-1}\mathrm{arcsinh}(As)\right)}
{\sqrt{1+{\lambda A^2 \over 4\pi^2 m^2}}\sqrt{1+A^2 t^2}\sqrt{1+A^2 s^2}}
-\frac{m A^2 t}{(1+A^2 t^2)}\delta(s-t)~,\nonumber\\
\zeta(t,s)&\equiv&
\frac{A^2 s\,\eta^{\prime\prime}\!\left(A^{-1}\mathrm{arcsinh}(At)-A^{-1}\mathrm{arcsinh}(As)\right)}
{\sqrt{1+{\lambda A^2 \over 4\pi^2 m^2}}\sqrt{1+A^2 t^2}({1+A^2 s^2})^{3/2}}
+\frac{m A^2 (1-A^2 t^2)}{(1+A^2 t^2)^2}\delta(s-t)~,\nonumber\\
\eta_{\perp}\!(t,s)&\equiv& \frac{\eta^{\prime\prime}\!\left(A^{-1}\mathrm{arcsinh}(At)-A^{-1}\mathrm{arcsinh}(As)\right)}
{\sqrt{1+{\lambda A^2 \over 4\pi^2 m^2}}(1+A^2 t^2)}+\frac{m A^2 t}{(1+A^2 t^2)}\delta(s-t)~,\\
\kappa(t,s)&\equiv&\kappa^{\prime\prime}\!\left(A^{-1}\mathrm{arcsinh}(At)-A^{-1}\mathrm{arcsinh}(As)\right)
~,\nonumber\\
\kappa_{\perp}\!(t,s)&\equiv& \frac{\kappa^{\prime\prime}\!\left(A^{-1}\mathrm{arcsinh}(At)-A^{-1}\mathrm{arcsinh}(As)\right)}
{\left(1+{\lambda A^2 \over 4\pi^2 m^2}\right)\sqrt{1+A^2 t^2}\sqrt{1+A^2 s^2}}~.\nonumber
\end{eqnarray}
Notice that, in contrast with the transverse fluctuations, the longitudinal one does not evolve according to a generalized Langevin equation: there is in (\ref{langevinlong}) an additional  dissipative force that depends (nonlocally) on $\delta x(t)$ itself. This distinction reflects the inherent anisotropy of the system in the inertial frame.

When we analyze the trajectory at quark proper times larger than the effective quark Compton wavelength, $\Delta\tau\gg z_m/\sqrt{1+A^2 z_m^2}$,
but over time intervals $\Delta t$ smaller than the characteristic time $A^{-1}$ set by the acceleration,
the low-frequency approximation of the previous subsection becomes appropriate, and the equations of motion again reduce to a local form,
\begin{equation} \label{locallangevinlong}
m_{\mathrm{th}} {d^2\delta x\over dt^2}(t)
+\eta_0(t){d\delta x\over dt}(t)
-\zeta_0(t)\delta x(t)
=\frac{\sqrt{1+{\lambda A^2 \over 4\pi^2 m^2}}f(t)}{\sqrt{1+A^2 t^2}}
~,
\end{equation}
and
\begin{equation} \label{locallangevinperp}
m_{\mathrm{th}} {d^2\delta \vec{x}_{\perp}\over dt^2}(t)
+\eta_{\perp 0}(t){d\delta \vec{x}_{\perp}\over dt}(t)
=\frac{\sqrt{1+{\lambda A^2 \over 4\pi^2 m^2}}\,\vec{f}_{\perp}(t)}{\sqrt{1+A^2 t^2}}
~,
\end{equation}
with $m_{\mathrm{th}}\equiv m\,m^{\prime\prime}_{\mathrm{th}}/m^{\prime\prime}=
m^{\prime\prime}_{\mathrm{th}}/\sqrt{1+\lambda A^2/ 4\pi^2 m^2}$,
\begin{equation}\label{f0}
\expec{f_i(t)}=0~,\qquad \expec{f_i(t)f_j(s)}=\delta_{ij}\kappa_{i 0}(t)\,\delta(t-s)~,
\end{equation}
where we have defined the time-dependent coefficients
\begin{eqnarray}\label{eta0kappa0}
\eta_0(t)&\equiv&\frac{A^2\left(\frac{\sqrt{\lambda} }{2\pi}\sqrt{1+A^2 t^2}-m t\right)}{(1+A^2 t^2)\sqrt{1+{\lambda A^2 \over 4\pi^2 m^2}}}~,\nonumber\\
\zeta_0(t)&\equiv&\frac{A^2\left(\frac{\sqrt{\lambda} }{2\pi}\sqrt{1+A^2 t^2}A^2 t+m (1-A^2 t^2)\right)}{(1+A^2 t^2)^2\sqrt{1+{\lambda A^2 \over 4\pi^2 m^2}}}~,\nonumber\\
\kappa_0(t)&\equiv& {\sqrt{\lambda}\over{2\pi^2}}A^3\sqrt{1+A^2 t^2}~,\\
\eta_{\perp 0}(t)&\equiv&\frac{A^2\left(\frac{\sqrt{\lambda} }{2\pi}(1+A^2 t^2)+\sqrt{1+{\lambda A^2 \over 4\pi^2 m^2}}\, m t\right)}{(1+A^2 t^2)^{3/2}\sqrt{1+{\lambda A^2 \over 4\pi^2 m^2}}}~,\nonumber\\
\kappa_{\perp 0}(t)&\equiv& {\sqrt{\lambda}\over{2\pi^2}}\frac{A^3}{\left(1+{\lambda A^2 \over 4\pi^2 m^2}\right)\sqrt{1+A^2 t^2}}~.\nonumber
\end{eqnarray}
These equations (as well as their nonlocal progenitors) reveal interesting structure in the fluctuation/dissipation setup induced by gluonic radiation in our strongly-coupled CFT. The main feature is the prominent time dependence of the problem, which is of course expected in the unprimed frame but in stark contrast with the situation in the doubly primed frame, where the quark is exposed to a static thermal medium. For the inertial observer, the effects produced by the emitted gluonic radiation do evolve with time, and the system can at best be regarded as quasi-stationary if it is examined in the regime $\Delta t\ll A^{-1}$.

In more detail, the velocity and the rate at which energy and longitudinal momentum are radiated by the quark follow from (\ref{quarktrajectory}), (\ref{radiationrate}) and (\ref{A})  as
\begin{eqnarray}\label{timedependence}
v&=&\frac{At}{\sqrt{1+A^2 t^2}}~, \nonumber\\
{d E_{\mbox{\scriptsize rad}}\over d t}&=& A^2\left(1-\frac{A^2 t}{\sqrt{{4\pi^2 m^2\over\lambda}+A^2}\sqrt{1+A^2 t^2}}\right)~,\\
{d p_{\mbox{\scriptsize rad}}\over d t}&=& A^2\left(\frac{At}{\sqrt{1+A^2 t^2}}-
\frac{A}{\sqrt{{4\pi^2 m^2\over\lambda}+A^2}}\right)~,
\end{eqnarray}
which clearly approach a constant only at very late (or very early) times, $|t|\gg A^{-1}$. In this limit, all of the dynamical coefficients (\ref{eta0kappa0}) are seen to vanish, except for the longitudinal force correlation coefficient $\kappa_0(t)$. The latter diverges, but does so at a rate that is too slow to give a finite contribution to the equation of motion (\ref{locallangevinlong}), which therefore becomes free.

It can be seen from (\ref{A}) that, as the external force on the quark increases from zero to its maximal value $F_{\mbox{\scriptsize{crit}}}=m^2/2\pi\sqrt{\lambda}$ (which would be strong enough to nucleate quark-antiquark pairs from the vacuum \cite{ctqhat,dragtime,damping}), the proper acceleration $A$ covers the full range $[0,\infty)$. Nevertheless, the situation of main physical interest is $\sqrt{\lambda} A/2\pi m < 1$, corresponding to a heavy (and therefore close to pointlike) quark that is not too violently accelerated (i.e., a quark that does not change velocity significantly in a time period smaller than its Compton wavelength $z_m$). Inspection of (\ref{eta0kappa0}) shows that under these circumstances $\zeta_0$ always starts out positive around $t=0$, becomes negative at a finite time and then asymptotically approaches zero from below as $t\to\infty$.  It is therefore particularly curious that the sign in front of the term linear $\delta x(t)$ in  (\ref{locallangevinlong})  (and (\ref{langevinlong})) turns out to be negative, because, when $\zeta_0>0$ ($\zeta>0$), this implies that the quark experiences a force that tends to push it further and further away from its average trajectory, instead of restoring it back towards equilibrium. Likewise, it is easy to see that $\eta_0$ (as well as $\eta$) generically becomes negative after a certain amount of time, implying that the effect of the emitted radiation is to speed up the longitudinal fluctuations of the quark instead of slowing them down. Of course, given that the equations of motion (\ref{langevinlong})-(\ref{eta0kappa0}) have been obtained simply by translating (\ref{langevinpp})-(\ref{locallangevinpp}) to the inertial frame, the combined effect of all terms appearing in them cannot possibly lead to runaway behavior.

We can similarly read off directly from \cite{rangamani} the expression for the displacement squared in the open Einstein frame and translate it to the inertial frame.\footnote{In the process one encounters an infrared divergence that happened to cancel out in the final result of \cite{rangamani}, and could be regularized as in \cite{deboer}.} We will refrain from writing out the results here, since they are not particularly illuminating. It is easy to extract from them the ballistic behavior expected at small times, but the diffusive regime of the doubly primed frame is not accessible to the inertial observer within the quasi-stationary regime.

\section*{Acknowledgements}

This work is dedicated to the memory of Blanca G\"uijosa. We are grateful to Alejandro Corichi, Roberto Emparan, Bartomeu Fiol, Antonio Garc\'{\i}a, Gast\'on Giribet, Dami\'an Hern\'andez, Mart\'\i n Kruczenski, David Mateos, \'Angel Paredes and Daniel Sudarsky for useful discussions. EC thanks the Theory Group at the University of
Texas at Austin for hospitality, and the Aspen Center for Physics for hospitality and partial
support. When this paper was begun, MCh was at the Facultad de Ciencias, UNAM, and his
work was supported by a DGAPA-UNAM postdoctoral fellowship. For the final stage of
this paper MCh is at the Departament de Fisica Fonamental, Universitat de
Barcelona, and his work is supported by a postdoctoral fellowship from Mexico's National Council of Science and Technology
(CONACyT).
The work of the  remaining authors was partially supported by CONACyT grants 50-155I, CB-2008-01-104649 and 50760, as well as DGAPA-UNAM grant IN116408 and by the National
Science Foundation under Grant No. PHY-0455649. EC and AG are members of CONACyT's High Energy Physics Network, whose support they gratefully acknowledge.


\begin{thebibliography}{99}

\bibitem{monizsharp}
  E.~J.~Moniz and D.~H.~Sharp,
  ``Absence of runaways and divergent self-mass in nonrelativistic quantum
  electrodynamics,''
  Phys.\ Rev.\  D {\bf 10} (1974) 1133;\\
  E.~J.~Moniz and D.~H.~Sharp,
  ``Radiation Reaction In Nonrelativistic Quantum Electrodynamics,''
  Phys.\ Rev.\  D {\bf 15} (1977) 2850.

\bibitem{martin}
  G.~D.~R.~Martin,
  ``Classical and Quantum Radiation Reaction,''
  arXiv:0805.0666 [gr-qc];\\
  A.~Higuchi and G.~D.~R.~Martin,
  ``Quantum Radiation Reaction and the Green's Function Decomposition,''
  Phys.\ Rev.\  D {\bf 74} (2006) 125002
  [arXiv:gr-qc/0608028];\\
  A.~Higuchi and G.~D.~R.~Martin,
  ``Radiation reaction on charged particles in three-dimensional motion in
  classical and quantum electrodynamics,''
  Phys.\ Rev.\  D {\bf 73} (2006) 025019
  [arXiv:quant-ph/0510043];\\
  A.~Higuchi and G.~D.~R.~Martin,
  ``The Lorentz-Dirac force from QED for linear acceleration,''
  Phys.\ Rev.\  D {\bf 70} (2004) 081701
  [arXiv:quant-ph/0407162];\\
  A.~Higuchi,
  ``Radiation reaction in quantum field theory,''
  Phys.\ Rev.\  D {\bf 66} (2002) 105004
  [Erratum-ibid.\  D {\bf 69} (2004) 129903]
  [arXiv:quant-ph/0208017].

\bibitem{rosenfelder}
  R.~Rosenfelder and A.~W.~Schreiber,
  ``An Abraham-Lorentz-like equation for the electron from the worldline
  variational approach to QED,''
  Eur.\ Phys.\ J.\  C {\bf 37}, 161 (2004)
  [arXiv:hep-th/0406062].

\bibitem{johnson}
  P.~R.~Johnson and B.~L.~Hu,
  ``Uniformly accelerated charge in a quantum field: From radiation  reaction
  to Unruh effect,''
  Found.\ Phys.\  {\bf 35} (2005) 1117
  [arXiv:gr-qc/0501029];\\
  ``Stochastic theory of relativistic particles moving in a quantum field.  II:
  Scalar Abraham-Lorentz-Dirac-Langevin equation, radiation reaction  and
  vacuum fluctuations,''
  Phys.\ Rev.\  D {\bf 65} (2002) 065015
  [arXiv:quant-ph/0101001];\\
  ``Stochastic theory of relativistic particles moving in a quantum field.  I:
  Influence functional and Langevin equation,''
  arXiv:quant-ph/0012137;\\
  ``Worldline influence functional: Abraham-Lorentz-Dirac-Langevin equation
  from QED,''
  arXiv:quant-ph/0012135.

  \bibitem{parentani}
  R.~Parentani,
  ``The Recoils of the accelerated detector and the decoherence of its
  fluxes,''
  Nucl.\ Phys.\  B {\bf 454} (1995) 227
  [arXiv:gr-qc/9502030].

 \bibitem{malda}
  J.~M.~Maldacena,
  ``The large $N$ limit of superconformal field theories and
  supergravity,''
  Adv.\ Theor.\ Math.\ Phys.\  {\bf 2}, 231 (1998)
  [Int.\ J.\ Theor.\ Phys.\  {\bf 38}, 1113 (1999)]
  [arXiv:hep-th/9711200].

\bibitem{gkp}
  S.~S.~Gubser, I.~R.~Klebanov and A.~M.~Polyakov,
  ``Gauge theory correlators from non-critical string theory,''
  Phys.\ Lett.\ B {\bf 428}, 105 (1998)
  [arXiv:hep-th/9802109].

\bibitem{w}
  E.~Witten,
  ``Anti-de Sitter space and holography,''
  Adv.\ Theor.\ Math.\ Phys.\  {\bf 2}, 253 (1998)
  [arXiv:hep-th/9802150].

 \bibitem{magoo}
  O.~Aharony, S.~S.~Gubser, J.~M.~Maldacena, H.~Ooguri and Y.~Oz,
   ``Large $N$ field theories, string theory and gravity,''
  Phys.\ Rept.\  {\bf 323}, 183 (2000)
  [arXiv:hep-th/9905111].

\bibitem{lorentzdirac}
  M.~Chernicoff, J.~A.~Garc\'\i a and A.~G\"uijosa,
  ``Generalized Lorentz-Dirac Equation for a Strongly-Coupled Gauge Theory,''
 Phys.\ Rev.\ Lett.\  {\bf 102} (2009) 241601
  [arXiv:0903.2047 [hep-th]].

  \bibitem{damping}
  M.~Chernicoff, J.~A.~Garc\'\i a and A.~G\"uijosa,
  ``A Tail of a Quark in $\cN=4$ SYM,''
  JHEP {\bf 0909} (2009) 080
  [arXiv:0906.1592 [hep-th]].

\bibitem{dirac}
  P.~A.~M.~Dirac,
  ``Classical theory of radiating electrons,''
  Proc.\ Roy.\ Soc.\ Lond.\  A {\bf 167} (1938) 148.

  \bibitem{hkkky}
  C.~P.~Herzog, A.~Karch, P.~Kovtun, C.~Kozcaz and L.~G.~Yaffe,
  ``Energy loss of a heavy quark moving through $\cN = 4$ supersymmetric
  Yang-Mills plasma,''
  JHEP {\bf 0607} (2006) 013
  [arXiv:hep-th/0605158].

\bibitem{gubser}
  S.~S.~Gubser,
  ``Drag force in AdS/CFT,''
  Phys.\ Rev.\  D {\bf 74} (2006) 126005
  [arXiv:hep-th/0605182].

\bibitem{gubserqhat}
  S.~S.~Gubser,
  ``Momentum fluctuations of heavy quarks in the gauge-string duality,''
  Nucl.\ Phys.\  B {\bf 790} (2008) 175
  [arXiv:hep-th/0612143].

\bibitem{ctqhat}
   J.~Casalderrey-Solana and D.~Teaney,
  ``Transverse momentum broadening of a fast quark in a $\cN = 4$ Yang Mills
  plasma,''
  JHEP {\bf 0704} (2007) 039
  [arXiv:hep-th/0701123].

\bibitem{dragtime}
  M.~Chernicoff and A.~G\"uijosa,
  ``Acceleration, Energy Loss and Screening in Strongly-Coupled Gauge
  Theories,''
  JHEP {\bf 0806}, 005 (2008)
  [arXiv:0803.3070 [hep-th]].

\bibitem{dominguez}
  F.~Dom\'\i nguez, C.~Marquet, A.~H.~Mueller, B.~Wu and B.~W.~Xiao,
  ``Comparing energy loss and $p_{\perp}$-broadening in perturbative QCD with
  strong coupling $\mathcal{N}=4$ SYM theory,''
  Nucl.\ Phys.\  A {\bf 811} (2008) 197
  [arXiv:0803.3234 [nucl-th]].

\bibitem{xiao}
  B.~W.~Xiao,
  ``On the exact solution of the accelerating string in $AdS_5$ space,''
  Phys.\ Lett.\  B {\bf 665} (2008) 173
  [arXiv:0804.1343 [hep-th]].

\bibitem{beuf}
  G.~Beuf, C.~Marquet and B.~W.~Xiao,
  ``Heavy-quark energy loss and thermalization in a strongly coupled SYM
  plasma,''
  arXiv:0812.1051 [hep-ph].

  \bibitem{nolineonthehorizon}
   A.~G\"uijosa and E.~J.~Pulido,
  in preparation.

\bibitem{mikhailov}
  A.~Mikhailov,
  ``Nonlinear waves in AdS/CFT correspondence,''
  arXiv:hep-th/0305196.

\bibitem{ppz}
  A.~Paredes, K.~Peeters and M.~Zamaklar,
  ``Temperature versus acceleration: the Unruh effect for holographic models,''
  JHEP {\bf 0904} (2009) 015
  [arXiv:0812.0981 [hep-th]].

\bibitem{dorn}
  H.~Dorn and H.~J.~Otto,
  ``Q anti-Q potential from AdS-CFT relation at $T \ge 0$: Dependence on
  orientation in internal space and higher curvature corrections,''
  JHEP {\bf 9809} (1998) 021
  [arXiv:hep-th/9807093].

  \bibitem{greensite}
  J.~Greensite and P.~Olesen,
  ``Worldsheet fluctuations and the heavy quark potential in the AdS/CFT
  approach,''
  JHEP {\bf 9904} (1999) 001
  [arXiv:hep-th/9901057].

  \bibitem{johnson2}
  P.~Johnson,
  ``Relativistic Particle Trajectories from Worldline Path Integral
  Quantization,''
{\it In the Proceedings of IEEE Particle Accelerator Conference (PAC 2001), Chicago, Illinois, 18-22 Jun 2001, pp 1781-1783}.

\bibitem{rangamani}
  J.~de Boer, V.~E.~Hubeny, M.~Rangamani and M.~Shigemori,
  ``Brownian motion in AdS/CFT,''
  JHEP {\bf 0907}, 094 (2009)
  [arXiv:0812.5112 [hep-th]].

\bibitem{sonteaney}
  D.~T.~Son and D.~Teaney,
  ``Thermal Noise and Stochastic Strings in AdS/CFT,''
  JHEP {\bf 0907}, 021 (2009)
  [arXiv:0901.2338 [hep-th]].

\bibitem{lm}
  A.~E.~Lawrence and E.~J.~Martinec,
  ``Black Hole Evaporation Along Macroscopic Strings,''
  Phys.\ Rev.\  D {\bf 50} (1994) 2680
  [arXiv:hep-th/9312127].

\bibitem{ff}
  V.~P.~Frolov and D.~Fursaev,
  ``Mining energy from a black hole by strings,''
  Phys.\ Rev.\  D {\bf 63}, 124010 (2001)
  [arXiv:hep-th/0012260].

\bibitem{maldaeternal}
  J.~M.~Maldacena,
  ``Eternal black holes in Anti-de-Sitter,''
  JHEP {\bf 0304}, 021 (2003)
  [arXiv:hep-th/0106112].

\bibitem{herzogson}
  C.~P.~Herzog and D.~T.~Son,
  ``Schwinger-Keldysh propagators from AdS/CFT correspondence,''
  JHEP {\bf 0303}, 046 (2003)
  [arXiv:hep-th/0212072].

\bibitem{ct}
  J.~Casalderrey-Solana and D.~Teaney,
  ``Heavy quark diffusion in strongly coupled $\cN = 4$ Yang Mills,''
  Phys.\ Rev.\  D {\bf 74} (2006) 085012
  [arXiv:hep-ph/0605199].

\bibitem{iancu}
  G.~C.~Giecold, E.~Iancu and A.~H.~Mueller,
  ``Stochastic trailing string and Langevin dynamics from AdS/CFT,''
  JHEP {\bf 0907}, 033 (2009)
  [arXiv:0903.1840 [hep-th]].

\bibitem{giecold}
  G.~C.~Giecold,
  ``Heavy quark in an expanding plasma in AdS/CFT,''
  JHEP {\bf 0906} (2009) 002
  [arXiv:0904.1874 [hep-th]].

\bibitem{casalderrey}
  J.~Casalderrey-Solana, K.~Y.~Kim and D.~Teaney,
  ``Stochastic String Motion Above and Below the World Sheet Horizon,''
  JHEP {\bf 0912}, 066 (2009)
  [arXiv:0908.1470 [hep-th]].

\bibitem{deboer}
  A.~N.~Atmaja, J.~de Boer and M.~Shigemori,
  ``Holographic Brownian Motion and Time Scales in Strongly Coupled Plasmas,''
  arXiv:1002.2429 [hep-th].

  \bibitem{unruh}
  W.~G.~Unruh,
   ``Notes on black hole evaporation,''
   Phys.\ Rev.\  D {\bf 14} (1976) 870;\\
  P.~C.~W.~Davies,
  ``Scalar particle production in Schwarzschild and Rindler metrics,''
  J.\ Phys.\ A  {\bf 8}, 609 (1975).

\bibitem{unruhreviews}
  L.~C.~B.~Crispino, A.~Higuchi and G.~E.~A.~Matsas,
  ``The Unruh effect and its applications,''
  Rev.\ Mod.\ Phys.\  {\bf 80}, 787 (2008)
  [arXiv:0710.5373 [gr-qc]];\\
  S.~Takagi,
  ``Vacuum Noise And Stress Induced By Uniform Acceleration: Hawking-Unruh
  Effect In Rindler Manifold Of Arbitrary Dimensions,''
  Prog.\ Theor.\ Phys.\ Suppl.\  {\bf 88}, 1 (1986).

\bibitem{hirayama}
  T.~Hirayama, P.~W.~Kao, S.~Kawamoto and F.~L.~Lin,
  ``Unruh effect and Holography,''
  arXiv:1001.1289 [hep-th].

  \bibitem{theisen}
  C.~Imbimbo, A.~Schwimmer, S.~Theisen and S.~Yankielowicz,
  ``Diffeomorphisms and holographic anomalies,''
  Class.\ Quant.\ Grav.\  {\bf 17} (2000) 1129
  [arXiv:hep-th/9910267].

\bibitem{roberto}
  R.~Emparan,
  ``AdS/CFT duals of topological black holes and the entropy of  zero-energy
  states,''
  JHEP {\bf 9906}, 036 (1999)
  [arXiv:hep-th/9906040].

\bibitem{bunch}
  T.~S.~Bunch,
  ``Stress Tensor Of Massless Conformal Quantum Fields In Hyperbolic
  Universes,''
  Phys.\ Rev.\  D {\bf 18}, 1844 (1978).

\bibitem{cdowker}
  P.~Candelas and J.~S.~Dowker,
  ``Field Theories On Conformally Related Space-Times: Some Global
  Considerations,''
  Phys.\ Rev.\  D {\bf 19}, 2902 (1979).

  \bibitem{bd}
  N.~D.~Birrell and P.~C.~W.~Davies,
  ``Quantum Fields In Curved Space,''
{\it  Cambridge Univ. Pr., UK (1982) 340p};\\
  R.~M.~Wald,
  ``Quantum field theory in curved space-time and black hole thermodynamics,''
{\it  Chicago Univ. Pr., USA (1994) 205 p}.

  \bibitem{hs}
  M.~Henningson and K.~Skenderis,
  ``The holographic Weyl anomaly,''
  JHEP {\bf 9807} (1998) 023
  [arXiv:hep-th/9806087].



\bibitem{uvir}
  L.~Susskind and E.~Witten,
  ``The Holographic Bound In Anti-De Sitter Space,''
  arXiv:hep-th/9805114;\\
  A.~W.~Peet and J.~Polchinski,
  ``UV/IR relations in AdS dynamics,''
  Phys.\ Rev.\ D {\bf 59} (1999) 065011
  [arXiv:hep-th/9809022].

\bibitem{bklt}
  V.~Balasubramanian, P.~Kraus, A.~E.~Lawrence and S.~P.~Trivedi,
  ``Holographic probes of anti-de Sitter space-times,''
  Phys.\ Rev.\  D {\bf 59} (1999) 104021
  [arXiv:hep-th/9808017].

\bibitem{fg}
C.~Fefferman, C.~R.~Graham,
 Conformal invariants,
in \emph{\'Elie Cartan et les Math\'ematiques d'Aujourd'hui},
(Ast\'erisque, 1985), 95.

\bibitem{dhss}
  S.~de Haro, S.~N.~Solodukhin and K.~Skenderis,
  ``Holographic reconstruction of spacetime and renormalization in the  AdS/CFT
  correspondence,''
  Commun.\ Math.\ Phys.\  {\bf 217} (2001) 595
  [arXiv:hep-th/0002230];\\
  K.~Skenderis,
  ``Asymptotically anti-de Sitter spacetimes and their stress energy  tensor,''
  Int.\ J.\ Mod.\ Phys.\  A {\bf 16}, 740 (2001)
  [arXiv:hep-th/0010138].

  \bibitem{skenderis}
  K.~Skenderis,
  ``Lecture notes on holographic renormalization,''
  Class.\ Quant.\ Grav.\  {\bf 19} (2002) 5849
  [arXiv:hep-th/0209067].

  \bibitem{bk}
  V.~Balasubramanian and P.~Kraus,
  ``A stress tensor for anti-de Sitter gravity,''
  Commun.\ Math.\ Phys.\  {\bf 208} (1999) 413
  [arXiv:hep-th/9902121].

  \bibitem{myers}
  R.~C.~Myers,
  ``Stress tensors and Casimir energies in the AdS/CFT correspondence,''
  Phys.\ Rev.\  D {\bf 60} (1999) 046002
  [arXiv:hep-th/9903203].

  \bibitem{ejm}
  R.~Emparan, C.~V.~Johnson and R.~C.~Myers,
  ``Surface terms as counterterms in the AdS/CFT correspondence,''
  Phys.\ Rev.\  D {\bf 60}, 104001 (1999)
  [arXiv:hep-th/9903238].

\bibitem{kk}
  A.~Karch and E.~Katz,
  ``Adding flavor to AdS/CFT,''
  JHEP {\bf 0206} (2002) 043
  [arXiv:hep-th/0205236].

\bibitem{martinfsq}
  J.~L.~Hovdebo, M.~Kruczenski, D.~Mateos, R.~C.~Myers and D.~J.~Winters,
  ``Holographic mesons: Adding flavor to the AdS/CFT duality,''
  Int.\ J.\ Mod.\ Phys.\  A {\bf 20} (2005) 3428.

\bibitem{acny}
  A.~Abouelsaood, C.~G.~Callan, C.~R.~Nappi and S.~A.~Yost,
  ``Open Strings In Background Gauge Fields,''
  Nucl.\ Phys.\  B {\bf 280} (1987) 599.

\bibitem{kundu}
  T.~Albash, V.~G.~Filev, C.~V.~Johnson and A.~Kundu,
  ``Quarks in an External Electric Field in Finite Temperature Large $N$ Gauge
  Theory,''
  JHEP {\bf 0808} (2008) 092
  [arXiv:0709.1554 [hep-th]].

\bibitem{bisognano}
  J.~J.~Bisognano and E.~H.~Wichmann,
  ``On The Duality Condition For A Hermitian Scalar Field,''
  J.\ Math.\ Phys.\  {\bf 16} (1975) 985;\\
  J.~J.~Bisognano and E.~H.~Wichmann,
  ``On The Duality Condition For Quantum Fields,''
  J.\ Math.\ Phys.\  {\bf 17} (1976) 303.
 G.~L.~Sewell,
  ``Quantum Fields on Manifolds: PCT and Gravitationally Induced Thermal States,''
  Ann.\ Phys.\  {\bf 141} (1982) 201.

\bibitem{cdeutsch}
  P.~Candelas and D.~Deutsch,
  ``On The Vacuum Stress Induced By Uniform Acceleration Or Supporting The
  Ether,''
  Proc.\ Roy.\ Soc.\ Lond.\  A {\bf 354} (1977) 79;\\
  ``Fermion Fields In Accelerated States,''
  Proc.\ Roy.\ Soc.\ Lond.\  A {\bf 362} (1978) 251.

\bibitem{deser}
  S.~Deser and O.~Levin,
  ``Accelerated detectors and temperature in (anti) de Sitter spaces,''
  Class.\ Quant.\ Grav.\  {\bf 14}, L163 (1997)
  [arXiv:gr-qc/9706018];\\
  S.~Deser and O.~Levin,
  ``Equivalence of Hawking and Unruh temperatures through flat space
  embeddings,''
  Class.\ Quant.\ Grav.\  {\bf 15}, L85 (1998)
  [arXiv:hep-th/9806223];\\
    S.~Deser and O.~Levin,
  ``Mapping Hawking into Unruh thermal properties,''
  Phys.\ Rev.\  D {\bf 59}, 064004 (1999)
  [arXiv:hep-th/9809159].

\bibitem{btz}
  M.~Ba\~nados, C.~Teitelboim and J.~Zanelli,
  ``The Black hole in three-dimensional space-time,''
  Phys.\ Rev.\ Lett.\  {\bf 69} (1992) 1849
  [arXiv:hep-th/9204099].

  \bibitem{bhtz}
  M.~Ba\~nados, M.~Henneaux, C.~Teitelboim and J.~Zanelli,
  ``Geometry of the (2+1) black hole,''
  Phys.\ Rev.\  D {\bf 48} (1993) 1506
  [arXiv:gr-qc/9302012].

\bibitem{duff}
  D.~M.~Capper and M.~J.~Duff,
  ``Trace anomalies in dimensional regularization,''
  Nuovo Cim.\  A {\bf 23} (1974) 173;\\
  S.~Deser, M.~J.~Duff and C.~J.~Isham,
  ``Nonlocal Conformal Anomalies,''
  Nucl.\ Phys.\  B {\bf 111} (1976) 45;\\
  S.~Deser and A.~Schwimmer,
  ``Geometric classification of conformal anomalies in arbitrary dimensions,''
  Phys.\ Lett.\  B {\bf 309} (1993) 279
  [arXiv:hep-th/9302047].

\bibitem{duffreview}
  M.~J.~Duff,
  ``Twenty years of the Weyl anomaly,''
  Class.\ Quant.\ Grav.\  {\bf 11}, 1387 (1994)
  [arXiv:hep-th/9308075].

\bibitem{cc}
  L.~S.~Brown and J.~P.~Cassidy,
  ``Stress Tensors And Their Trace Anomalies In Conformally Flat Space-Times,''
  Phys.\ Rev.\  D {\bf 16} (1977) 1712;\\
  T.~S.~Bunch,
  ``On Renormalization Of The Quantum Stress Tensor In Curved Space-Time By
  Dimensional Regularization,''
  J.\ Phys.\ A  {\bf 12}, 517 (1979);\\
  A.~Cappelli and A.~Coste,
  ``On the stress tensor of conformal field theories in higher dimensions,''
  Nucl.\ Phys.\  B {\bf 314} (1989) 707.


\end{thebibliography}
\end{document}